\documentclass[11pt]{article}
\usepackage{amsmath}
\usepackage{amsfonts}
\usepackage{amssymb, amsmath, cite}

\newtheorem{theorem}{Theorem}
\newtheorem{lemma}{Lemma}

\newtheorem{remark}{Remark}

\newcommand\be{\begin{equation}}
\newcommand\ee{\end{equation}}
\newcommand\ber{\begin{eqnarray}}
\newcommand\eer{\end{eqnarray}}
\newcommand\berr{\begin{eqnarray*}}
\newcommand\eerr{\end{eqnarray*}}
 \newcommand\re{\mathrm{e}}
\newcommand{\ud}{\mathrm{d}}

\newcommand{\ito}{\int_{\Omega}}
\newcommand{\vep}{\varepsilon}
{\setlength\arraycolsep{2pt}

\begin{document}

\title{ \bf\Large  The Existence  of Multi-vortices for  a Generalized Self-dual Chern--Simons Model \footnote{This work was supported by the National
Natural Science Foundation of China under grant 11026077 and  Henan
Basic Science and Frontier Technology Program Funds under grant
112300410054.}}
\author{Xiaosen Han\footnote{Email: xiaosenhan@gmail.com} \\[2mm]
  {\it\small  Institute of Contemporary Mathematics, Henan University, Kaifeng 475004, China  }\\[2mm]
  {\it\small School of Mathematics and Information Science, Henan University, Kaifeng 475004, China}}

\date{}

\maketitle

\begin{quote}{
{{\bfseries Abstract.}  In this paper we establish the existence of
multi-vortices for  a generalized self-dual Chern--Simons model.
Doubly periodic vortices, topological and non-topological vortex
solutions are constructed for this  model. For the existence of
doubly periodic vortex solutions, we establish an explicitly
necessary and sufficient condition. It is difficult  to get
topological multi-vortex solutions due to the non-canonical
structure of the equations.  We overcome this difficulty by
constructing a suitable sub-solution for the reduced equation. This
technique maybe applied to the problems with  similar structures.
For the existence of non-topological solutions we use a shooting
argument.}}



\end{quote}
\section{Introduction}\label{s1}

In mathematical physics  static solutions to   gauge field equations
with broken symmetry  in two-space dimensions are often called
vortices. Magnetic vortices play important roles in many areas of
theoretical physics including superconductivity
\cite{abri,jata,gila}, electroweak
theory\cite{amol1,amol2,amol3,amol4}, and
cosmology\cite{hiki,kibb,vish}.  The first and also the best-known
rigorous mathematical  construction of magnetic vortices was due to
Taubes\cite{jata,taub1,taub2} regarding the existence and uniqueness
of static solutions of the Abelian Higgs model or the
Ginzburg--Landau model\cite{gila}. Since then there have been much
mathematical work about the existence and properties of such
vortices. See, for example, the references \cite{
yang1,tara4,bch1,nwu,dgp1,spya1,spya2,bebrhe,wein,beri,lfh1,lfh2,ovsi,bpt1,serf,pari,msz}.
It is also natural to consider  the dyon-like vortices, often
referred to electrically charged magnetic vortices, carrying both
magnetic and electric charges. Such dually charged vortices  are
very useful in several issues in theoretical physics such as
high-temperature superconductivity \cite{khfr,mnk}, the
Bose-Einstein condensates \cite{igrc,kaoh}, optics\cite{bec}, and
the quantum Hall effect \cite{soko}.

It is now  well-known that there is no finite--energy dually charged
vortices in two--space dimensions for the classical
Yang--Mills--Higgs equations, Abelian or non-Abelian. This is known
as  the Julia--Zee theorem \cite{juze}, whose rigorous mathematical
proof was carried out in \cite{spya5}. To accommodate dually charged
vortices,  some effort has been done in
\cite{jate1,schon,djt1,djt2,vesch1,vesch2,kukh1,pakh1}, where the
Chern--Simons terms are introduced into the action Lagrangian.
However, the full Chern--Simons--Higgs equations are very difficult
to tackle. Only radial case has been solved in \cite{chgsya}.  Since
the work of Hong, Kim, and Pac\cite{hkp} and Jackiw and
Weinberg\cite{jw1} was published, there has been some rigorous
mathematical results on the existence of  electrically charged
vortices. In \cite{hkp,jw1}, the Yang--Mills(or Maxwell) term is
removed from the action Lagrangian density while the Chern--Simons
term alone governs electromagnetism, which is physically sensible at
large distance and low energies.  When the Higgs potential takes a
sextic form, the static equations of motion can be reduced to a
Bogolmol'ny type system of first order equations \cite{bogo}, which
enables one to make rigorous mathematical studies of such solutions.
In such a setting, topological multi-vortices with quantized
charges\cite{wangr,spya4}, non-topological multi-vortices with
fractional values of charges \cite{spya3,chmy,chim,cfl} and periodic
vortices \cite{caya1,taran96,cho1,djlw,linyan,tara5,nota1} are all
present.

  Recently, a generalized Chern--Simons model was proposed in \cite{bhsm}.
  With a non-canonical kinetic term  for the complex scalar field and a special choice of the Higgs potential,
     a generalized self-dual Chern--Simons equation can be
  obtained. Despite  some numerical work  done in \cite{bhsm}, up to now, there is no rigorous mathematical
  analysis for this model.

 Our purpose of this paper is to carry out a rigorous mathematical  analysis of
 this generalized self-dual Chern--Simons model. Specifically,  we will
 establish the existence of doubly  periodic multi-vortices, topological and
 non-topological vortex solutions to the generalized  self-dual Chern--Simons  model.
 We reduce the problem to a semi-linear elliptic equations
 with Dirac source terms characterizing  the locations of the vortices.
  To get the existence of vortices over a doubly  periodic domain, we apply
 the method developed by Caffarelli and Yang\cite{caya1}  to construct   suitable sub-solutions
for the reduced equation. However, it is difficult to get the
existence of  topological solutions, due to the non-canonical
structure of the reduced equations.   To solve this difficulty, we
 find  a technical  sub-solution, which is useful for the problems of similar structures. For the existence of
non-topological solution,  we  apply  a shooting argument, which was
used in \cite{chmy,yang1}.

 The rest of our paper is organized as follows. In section 2,
we formulate our problem  and state our main results. In section 3
we prove the existence of periodic vortices. In section 4 and
section 5 we give the proofs of topological solutions and
non-topological solutions, respectively.

\section{Generalized Chern--Simons vortices}\label{s2}

\setcounter{equation}{0}\setcounter{remark}{0}
  In this section we   derive the
  generalized  self-dual Chern--Simons equations, while in \cite{bhsm} only radial case is considered.
  We  adapt the notation in \cite{jw1}. The $(2+1)$-dimensional Minkowski
  space metric tensor $g_{\mu\nu}$ is $\mathrm{diag}(1,-1,-1)$, which is used to raise and  lower indices. The
  Lagrangian action  density  of the Chern--Simons-Higgs theory is
  given by the expression
  \be
  \mathcal {L}_s=\frac\kappa4\vep^{\alpha\beta\gamma}A_\alpha
  F_{\beta\gamma}+D_\mu\phi \overline{D^\mu\phi}-V(|\phi|),\label{2.1}
  \ee
  where $D_\mu=\partial_\mu-iA_\mu$ is the gauge-covariant derivative,  $A_\mu \, (\mu=0,1,2)$ is 3-vector field
  called the  Abelian gauge field, $\phi$ is  a complex scalar field
  called the Higgs field, $F_{\alpha\beta}=\partial_\alpha A_\beta-\partial_\beta
  A_\alpha,$ is the induced electromagnetic field,  $\alpha, \beta, \mu,\nu=0, 1, 2$, $\kappa>0$ is a
  constant referred to as the Chern--Simons
  coupling parameter, $\vep^{\alpha\beta\gamma}$ the Levi-Civita totally
  skew-symmetric tensor with $\vep^{012}=1$,  $V$ is the Higgs
  potential function, and the summation convention over repeated indices
  is observed.

 In \cite{bhsm},  by changing the kinetic term  of the Higgs field in the model,  \eqref{2.1}  is
  modified into a non-canonical form, the      new
  Lagrangian density reads
  \be
  \mathcal {L}_G=\frac\kappa4\vep^{\alpha\beta\gamma}A_\alpha F_{\beta\gamma}+w(|\phi|)D_\mu\phi \overline{D^\mu\phi}-V(|\phi|),\label{2.2}
  \ee
where $w(|\phi|)$ is a function of the Higgs field.

  The Euler--Lagrangian equations associated with the action density
  $\mathcal{L}_G$ \eqref{2.2} are
   \ber
   \frac\kappa2\vep^{\alpha\beta\gamma}A_\alpha   F_{\beta\gamma}=-j^\alpha=iw(|\phi|)(\phi\overline{D^\alpha\phi}-\overline{\phi} D^\alpha\phi), \label{2.3}\\
    wD_\mu D^\mu\phi+\partial_\mu wD^\mu\phi+\frac{\partial V}{\partial\overline{\phi}}-|D_\mu \phi|^2\frac{\partial w}{\partial \overline{\phi}}=0, \label{2.4}
   \eer
  where $j^\alpha=(\rho, \mathbf{j})$ is the current density and
  $F_{12}$ is the magnetic field.

 We are interested in the static solutions of the equations
 \eqref{2.3} and \eqref{2.4} over a doubly periodic domain $\Omega$  such that the
field configurations are subject to the 't Hooft boundary condition
\cite{hoof,waya,yang1} under which periodicity is achieved modulo
gauge transformations, and over  the full plane $\mathbb{R}^2$. The
$\alpha=0$ component of the
 equation \eqref{2.3} reads
  \be
  \kappa F_{12}=j^0=\rho=-2A_0|\phi|^2w(\phi),\label{2.5}
  \ee
 the magnetic flux $\Phi$ and the electric charge $Q$ are related by
 the formula
 \be
  \Phi= \int F_{12}\ud x=\frac1\kappa \int \rho\ud x=\frac{Q}{\kappa}.\label{2.6}
  \ee
  Here and in the following,    our integration is always conducted over the doubly periodic
  domain $\Omega$
  or the full plane $\mathbb{R}^2$.
 The energy density is then given by
  \be
  \mathcal {E}=\frac{\kappa^2}{2}\frac{F^2_{12}}{|\phi|^2w}+w|D_j \phi|^2+V(|\phi|).\label{2.7}
  \ee
  As in \cite{bhsm} we choose  special forms  for $w(|\phi|)$ and the Higgs potential
  function $V(\phi)$ as follows
   \[w(|\phi|)=3(1-|\phi|^2)^2, \quad V(|\phi|)=\frac{3}{\kappa^2}|\phi|^2(1-|\phi|^2)^8.\]

 Then   the  energy density \eqref{2.7} implies that   a finite-energy solution
 of the solution to \eqref{2.3}-\eqref{2.4} over $\mathbb{R}^2$ satisfies the condition
  \be
   |\phi(x)|\to 1\quad\text{as}\quad |x|\to +\infty\label{2.10}
  \ee
or
 \be
   |\phi(x)|\to 0\quad\text{as}\quad |x|\to +\infty,\label{2.11}
  \ee
where the former is called topological  and the latter is called
 non-topological, see \cite{yang1,tara4,dun}.

 The energy density  \eqref{2.7} can be rewritten  as
 \berr
  \mathcal {E}&=&\left[\frac{\kappa}{2\sqrt{3}}\frac{F_{12}}{|\phi|(1-|\phi|^2)}\mp \frac{\sqrt{3}}{\kappa}|\phi|(1-|\phi|^2)^4\right]^2
  +3(1-|\phi|^2)^2\left|D_1\phi\pm \mathrm{i}D_2\phi\right|^2\\
   &&\pm (1-|\phi|^2)^3F_{12}\pm\mathrm{i}3(1-|\phi|^2)^2\big(\overline{D_1\phi}D_2\phi-D_1\phi\overline{D_2\phi}\big).
  \eerr

 If $(\phi, A)$ is a finite-energy solution with winding number $N$ of
 \eqref{2.3}-\eqref{2.4}, following  \cite{jata,yang1}, we can show that
 \[
 \int F_{12}\ud x=2\pi N.
 \]

  Integrating over the doubly periodic domain $\Omega$ or the
 full plane $\mathbb{R}^2$, we have
  \ber
   E(\phi, A)&=&\int \mathcal {E}\ud x\nonumber\\
   &=&\int\left\{\left[\frac{\kappa}{2\sqrt{3}}\frac{F_{12}}{|\phi|(1-|\phi|^2)}\mp \frac{\sqrt{3}}{\kappa}|\phi|(1-|\phi|^2)^4\right]^2
   +3(1-|\phi|^2)^2\left|D_1\phi\pm \mathrm{i}D_2\phi\right|^2\right\}\ud x\nonumber\\
   &&\pm 2\pi N\pm\int \mathrm{Im}\Big\{\partial_j\vep_{jk}\left(\left[|\phi|^2-(1-|\phi|^2)^2-2\right]\bar{\phi}(D_k\phi)\right)\Big\}\ud x.\label{2.7'}
 \eer
 Therefore, we can get the following lower bound of the energy
 \[
  E(\phi, A)\ge \pm2\pi N.
 \]
 Then we see from \eqref{2.7'} that  such a lower bound is attained if and only if $(\phi, A)$
 satisfies the following self-dual or anti-self-dual system
 \ber
  D_1\phi-\mathrm{i}D_2\phi=0,\label{2.8}\\
  F_{12}+\frac{6}{\kappa^2}|\phi|^2(1-|\phi|^2)^5=0\label{2.9}
 \eer
 or
 \ber
  D_1\phi+\mathrm{i}D_2\phi=0,\label{2.12}\\
  F_{12}-\frac{6}{\kappa^2}|\phi|^2(1-|\phi|^2)^5=0.\label{2.13}
 \eer
 It is easy to check that if $(\phi, A)$ is a solution of the system
 \eqref{2.8}-\eqref{2.9},  then $(\bar{\phi}, -A)$ is the solution of \eqref{2.12}-\eqref{2.13}. In addition,
 in view of \eqref{2.5}, any  solution of  \eqref{2.8}-\eqref{2.9}
 or  \eqref{2.12}-\eqref{2.13}
 is also the solution of \eqref{2.3}-\eqref{2.4}. Consequently, in the sequel
 we only consider  \eqref{2.8}-\eqref{2.9}.

 To formulate our problem more properly, as in \cite{jata,caya1,
yang1} we can see that the zeros of $\phi$ are isolated with integer
multiplicities. These zeros  are often  referred to   vortices. Let
the zeroes of $\phi$ be $p_1, p_2, \dots, p_m$ with  multiplicities
$n_1, n_2, \dots, n_m$, respectively. Then,  $\sum_{i=1}^m n_i=N$
gives the winding number of the solution and the total vortex
number. We aim to look for $N$-vortex solutions of
\eqref{2.8}-\eqref{2.9} such that, $\phi$ has $m$ zeros, say $p_1,
p_2, \dots, p_m$ with multiplicities $n_1, n_2, \dots, n_m$,
respectively, and $\sum_{i=1}^m n_i=N$.

 For the generalized Chern--Simons equations \eqref{2.8}-\eqref{2.9}, we are interested in three situations.
 In the first situation  the equations \eqref{2.8}-\eqref{2.9} will
be studied over a doubly periodic domain $\Omega$ such that the
field configurations are subject to the 't Hooft boundary condition
\cite{hoof,waya,yang1} under which periodicity is achieved modulo
gauge transformations. In the second and the third situations  the
equations are studied over the full plane $\mathbb{R}^2$  under the
topological condition \eqref{2.10} and non-topological condition
\eqref{2.11}, respectively.

The  main results of this paper read as follows.

  \begin{theorem} \label{th1}{\bf(Existence of Doubly Periodic Vortices)}
   Let $p_1, p_2, \dots, p_m\in\Omega$, $n_1, n_2, \dots, n_m$ be
   some  positive integers and $N=\sum_{i=1}^mn_i$. There exists a critical
   value
   \[\kappa_c\in \left(0, \sqrt{\frac{5^5}{6^5}\frac{|\Omega|}{2\pi N}}\right]\]
 of the coupling parameter such that the self-dual equations
 \eqref{2.8}-\eqref{2.9} admit a solution $(\phi, A)$ for which $p_1,
 p_2,\dots,p_m$ are zeros of $\phi$ with multiplicities $n_1, n_2, \dots,
 n_m$, if and only if $0<\kappa\le \kappa_c$. When $0<\kappa\le
 \kappa_c$, the solution $(\phi, A)$ also satisfies the following
 properties.

   The energy, magnetic flux, and electric charge are given by
    \be
     E=2\pi N,\quad \Phi=2\pi N, \quad Q=2\kappa\pi N.\label{2.14}
    \ee

    The solution  $(\phi, A)$ can be chosen such that the magnitude
    of $\phi$, $|\phi|$ has the largest possible values.

    Let the prescribed data be denoted by $S=\{p_1, p_2,\dots p_m; n_1, n_2,
    \dots,n_m\}$, where $n_i$ may be zero for $i=1,\dots,m$, and
    denote the dependence of $\kappa_c$ on $S$ by $\kappa_c(S)$. For
    $S'=\{p_1, p_2,\dots p_m; n'_1, n'_2, \dots,n'_m\}$, we denote $S\le S'$
    if $n_1\le n_1', \dots, n_m\le n_m'$. Then $\kappa_c$ is a
    decreasing function of $S$ in the sense that
     \be \kappa_c(S)\ge\kappa_c(S'), \quad {if}\quad S\le S'.\label{2.15}\ee
  \end{theorem}

  \begin{theorem}\label{th2}{\bf(Multiple Existence of Doubly Periodic Vortices)}
     Let $p_1, p_2, \dots, p_m\in\Omega$, $n_1, n_2, \dots, n_m$ be
   some positive  integers and $N=\sum_{i=1}^mn_i$ and $\kappa_c$ be given in
   Theorem \ref{th1}. If  $0<\kappa<\kappa_c$, then, in addition to
   the maximal solution  $(\phi, A)$ given in Theorem \ref{th1}, the
   self-dual equations \eqref{2.8}-\eqref{2.9} have a second solution  $(\tilde{\phi}, \tilde{A})$
    satisfying \eqref{2.14} and for which $p_1, p_2, \dots, p_m$ are the zeros of $\tilde{\phi}$
    with multiplicities $n_1, n_2, \dots, n_m$.
  \end{theorem}

  \begin{theorem}\label{th3}{\bf(Topological Solution)}
     Let $p_1, p_2, \dots, p_m\in\mathbb{R}^2$, $n_1, n_2, \dots, n_m$ be
   some positive integers and $N=\sum_{i=1}^mn_i$. The self-dual
   equations \eqref{2.8}-\eqref{2.9} admit a topological solution $(\phi,
   A)$ such that the zeros of $\phi$ are exactly $p_1, p_2, \dots,
   p_m$ with corresponding multiplicities $n_1, n_2, \dots, n_m$. Moreover,
    the energy, magnetic flux, and the charges are all quantized
   \be
     E=2\pi N,\quad \Phi=2\pi N, \quad Q=2\kappa\pi N.\label{2.16}
    \ee
  The solution is maximal in the sense that the  Higgs field $\phi$
 has the largest possible magnitude among all the solutions with the
 same zero distribution and local vortex charges in the full plane.
  \end{theorem}

  \begin{theorem} \label{th4}{\bf (Radially Symmetric Topological    Solution)}
  For any point $\tilde{x}\in \mathbb{R}^2$ and a given integer
  $N\ge0$, the self-dual equations \eqref{2.8}-\eqref{2.9} admit a unique   topological
  solution $(\phi , A ), $ which is radially symmetric about the point $\tilde{x}$,  such that $\tilde{x}$ is the zero of
  $\phi$ with multiplicities $N$. Moreover, the the energy, magnetic flux, and the charges are all
  quantized, given by \eqref{2.16}.
  \end{theorem}

   \begin{theorem}  \label{th5}{\bf (Radially Symmetric Non-topological    Solution)}
    For any point $\tilde{x}\in \mathbb{R}^2$ and a given integer  $N\ge0$,
   then for all $\beta>2N+4$, the self-dual equations \eqref{2.8}-\eqref{2.9} allows a     non-topological
  solution $(\phi , A ), $ which is radially symmetric about the point $\tilde{x}$,  such that $\tilde{x}$ is the zero of
  $\phi$ with multiplicities $N$ and realizing the  prescribed decay
  properties,
     \be
     |\phi|^2=O(r^{-\beta}), \quad  |D_j\phi|^2=O(r^{-(2+\beta)}),\, j=1,2,\quad     F_{12}=O(r^{-\beta})\label{2.17}
     \ee
     for large $r=|x-\tilde{x}|>0$ and the corresponding values of
     energy, magnetic flux, electric charge are given by the formula
     \be
      E=2\pi N+\pi\beta,\quad \Phi=2\pi N+\pi\beta, \quad Q=
      \kappa(2\pi N+\pi\beta).\label{2.18}
     \ee

  \end{theorem}
 \begin{remark}
  For the  non-topological solution  with arbitrary  distributed vortices to  the generalized
  Chern--Simons  equations \eqref{2.8}-\eqref{2.9}, since it is more  involved, we will deal with it in a forthcoming paper.
 \end{remark}

\section{ Existence of doubly periodic vortices}\label{s3}

\setcounter{equation}{0} \setcounter{theorem}{0}

In this section we aim to establish the existence of   vortex
solution to the generalized Chern--Simons equations
\eqref{2.8}-\eqref{2.9} over a doubly periodic domain $\Omega$.  In
other words, we present the proofs of Theorem \ref{th1}-\ref{th2} in
this section.

For convenience, we reduce the self-dual equation
\eqref{2.8}-\eqref{2.9} to scalar nonlinear elliptic equation with
Dirac source terms.  To this end, we complexify the variables
 \[z=x^1+\mathrm{i}x^2,  \quad A=A_1+\mathrm{i}A_2.
 \]
Let
\[\partial=\frac12(\partial_1-\mathrm{i}\partial_2)\quad \bar{\partial}=\frac12(\partial_1+\mathrm{i}\partial_2)\]
 Then from \eqref{2.8}, we can get
 \be
  F_{12}=-2\partial\bar{\partial}\ln|\phi|^2=-\frac12\Delta\ln|\phi|^2.\label{2.12'}
  \ee
Inserting \eqref{2.12'} into \eqref{2.9} gives rise to  the reduced
equation
 \be
  \Delta\ln|\phi|^2=\lambda |\phi|^2(|\phi|^2-1)^5  \label{2.13'}
 \ee
away from the zeros of $\phi$, where we write \[
\lambda\equiv\frac{12}{\kappa^2}\] throughout this paper.

 Counting all  the multiplicities of the zeros of $\phi$, we write
 the prescribed zero set as $Z(\phi)=\{p_1, \dots, p_N\}$. Let $|\phi|^2=\mathrm{e}^u.$
   Then the generalized self-dual Chern--Simons equations
   \eqref{2.8}-\eqref{2.9} are transformed into the following
   scalar equation
   \be
    \Delta  u=\lambda\mathrm{e}^u(\mathrm{e}^u-1)^5+4\pi\sum\limits_{s=1}^N\delta_{p_s}, \quad \text{in}\quad \Omega,\label{2.14'}
   \ee
where $\delta_p$ is the Dirac distribution centered at $p\in\Omega$.

Conversely, if $u$ is a solution of \eqref{2.14'}, we can obtain a
solution of \eqref{2.8}-\eqref{2.9} according to the transformation
 \berr
  &&\phi(z)=\exp{\left(\frac12u(z)+\mathrm{i}\sum\limits_{s=1}^N\arg(z-p_s)\right)},\\
   &&A_1(z)=-2\mathrm{Re}\{\mathrm{i}\bar{\partial}\ln\phi\}, \quad A_2(z)=-2\mathrm{Im}\{\mathrm{i}\bar{\partial}\ln\phi\}.
  \eerr
Then it is sufficient to solve \eqref{2.14'}.

Let $u_0$ be a solution of the equation (see \cite{aubi})
 \be \Delta u_0=-\frac{4\pi N}{|\Omega|}+4\pi\sum\limits_{j=1}^{N}\delta_{p_j}. \label{b.2}
 \ee
 Setting $u=u_0+v$, the equation (\ref{2.14'}) can be reduced to the
 following equation
 \be
 \Delta v=\lambda\re^{u_0+v}(\re^{u_0+v}-1)^5+\frac{4\pi N}{|\Omega|}. \label{b.3}
 \ee

It is easy to check that the function $f(t)=\re^t(\re^t-1)^5 (t\in
\mathbb{R})$ has a unique minimal value $-\frac{5^5}{6^6}$. Then, if
$v$ is a solution of (\ref{b.3}), we have
  \be
  \Delta v\ge-\frac{5^5}{6^6}\lambda+\frac{4\pi N}{|\Omega|}.\label{b.4}
  \ee Integrating
(\ref{b.4}) over  $\Omega$, we have
\[0\ge -\frac{5^5}{6^6}\lambda|\Omega|+4\pi N,\]
i.e., \be
 \lambda\ge\frac{6^6}{5^5}\frac{4\pi N}{|\Omega|}, \label{b.5}
\ee which is a necessary condition for the existence of solutions to
(\ref{2.14'}).

 As in \cite{caya1} or  chapter 5 in \cite{yang1}  we can use  a super- and sub-solution method to establish the existence results for
\eqref{2.14'}.

To solve (\ref{b.3}), we introduce the following iterative scheme
  \ber\left\{\begin{array}{lll}
  (\Delta-K)v_n=\lambda \re^{u_0+v_{n-1}}(\re^{u_0+v_{n-1}}-1)^5-Kv_{n-1}+\frac{4\pi N}{|\Omega|},\\
  n=1,2,\dots,\\[1mm]
  v_0=-u_0,
\end{array}\label{b.6}
\right.
  \eer
where $K>0$ is a constant to be determined.

\begin{lemma}\label{lem1}
Let $\{v_n\}$ be  the sequence defined by \eqref{b.6} with
$K\ge6\lambda$. Then \be
 v_0>v_1>v_2>\cdots>v_n>\cdots>v_\_\label{b.7}
 \ee
for any sub-solution $v_\_$ of \eqref{b.3}. Therefore, if
\eqref{b.3} has a sub-solution,
 the sequence $\{v_n\}$ converge to a solution of \eqref{b.3} in the space $C^k(\Omega)$
 for any $k\ge0$ and such a solution is the maximal solution of the equation.
\end{lemma}

{\it Proof.} \quad We prove by \eqref{b.7} by induction.

  We prove the case $n=1$ first. From \eqref{b.6} we have,
  \[(\Delta-K) v_1=Ku_0+\frac{4\pi N}{|\Omega|}, \]
 which implies $v_1\in C^\infty(\Omega)\cap C^\alpha(\Omega)$ for some $0<\alpha<1$.
 Noting that $(\Delta-K)(v_1-v_0)=0 $ in  $\Omega-\{p_1, p_2, \dots, p_N\}$ and $v_1-v_0<0$ on $\partial\Omega_\vep$, where $\Omega_\vep$ is the complement of $\bigcup_{j=1}^N\{x|\,\, |x-p_j|<\vep\}$ in $\Omega$ for $\vep$ sufficiently small, using the maximum principle we have $v_1-v_0<0$ in $\Omega_\vep$. Therefore, we have $v_1-v_0<0$ in $\Omega$.

 Assume that $v_0>v_1>\cdots>v_k.$  From \eqref{b.6} and $K>6\lambda$ we obtain
 \berr
 (\Delta-K)(v_{k+1}-v_k)&=&\lambda[\re^{u_0+v_{k}}(\re^{u_0+v_{k}}-1)^5-\re^{u_0+v_{k-1}}(\re^{u_0+v_{k-1}}-1)^5]-K(v_k-v_{k-1})\\
  &=&[\lambda\re^{u_0+\xi}(\re^{u_0+\xi}-1)^4(6\re^{u_0+\xi}-1)-K](v_k-v_{k-1})\\
  &\ge&(5\lambda-K)(v_k-v_{k-1})\\
  &\ge&0,
 \eerr
 where $v_k\le \xi\le v_{k-1}$. Using maximum principle again, we have $v_{k+1}<v_k$ in $\Omega$.

 Now we prove the lower bound in \eqref{b.7} in terms of the sub-solution $v\_$ of \eqref{b.3}, that is,  $v\_\in C^2(\Omega)$ and
 \be
 \Delta v\_\ge \lambda\re^{u_0+v\_}(\re^{u_0+v\_}-1)^5+\frac{4\pi N}{|\Omega|}.\label{b.8}
 \ee
Noting that $v_0=-u_0$ and \eqref{b.8}, we have
\[\Delta (v\_-v_0)\ge\lambda\re^{u_0+v\_}(\re^{u_0+v\_}-1)^5=\lambda\re^{v\_-v_0}(\re^{v\_-v_0}-1)^5 \quad\text{in}\quad\Omega\setminus\{p_1,\dots, p_N\}.\]
 If $\vep>0$ is small, we see that $v\_-v_0<0$ on $\partial\Omega_\vep$. Then, by maximum principle, we obtain $v\_-v_0<0$ in $\Omega_\vep$. Therefore, $v\_-v_0<0$ throughout $\Omega$.

Now assume $v\_<v_k$ for some $k\ge0$. It follows from \eqref{b.7},
\eqref{b.8} and the fact  $K>6\lambda$ that
 \berr
  (\Delta-K) (v\_-v_{k+1})&\ge& \lambda[\re^{u_0+v\_}(\re^{u_0+v\_}-1)^5-\re^{u_0+v_{k}}(\re^{u_0+v_{k}}-1)^5]-K(v\_-v_k)\\
  &\ge&[\lambda\re^{u_0+\xi}(\re^{u_0+\xi}-1)^4(6\re^{u_0+\xi}\check{}-1)-K](v\_-v_k)\\
  &\ge&(5\lambda-K)(v\_-v_k)\\
  &\ge&0,
 \eerr
where   $v\_\le \xi\le v_k$.  Applying maximum principle again, we
get $v\_<v_{k+1}$.

The convergence of the sequence $\{v_n\}$ can be obtained by a
standard bootstrap argument. Then Lemma \ref{lem1} follows.

In what follows we just need to construct  a sub-solution of
\eqref{b.3}. Indeed, we have the following lemma.
\begin{lemma}\label{lem2}
 If $\lambda>0$ is sufficiently large,  the equation \eqref{b.3} admits a sub-solution  satisfying \eqref{b.8}.
\end{lemma}

{\it Proof.} \quad  Choose $\vep>0$ sufficiently small such that the
balls
 \[ B(p_j, 2\vep)=\{x\in \Omega| \quad |x-p_j|<2\vep\}, \,\, j=1, 2, \cdots, N\]
verify $B(p_i, 2\vep)\bigcap B(p_j, 2\vep)=\emptyset$ if $i\neq j$.
Let $f_\vep$ be a smooth function defined on $\Omega$ such that
$0\le f_\vep\le 1$ and
  \berr f_\vep=\left\{\begin{array}{lll}1,&x\in B(p_j, \vep), \,\,  j=1, 2, \cdots, N,\\
  0, &x\notin\bigcup\limits_{j=1}^N B(p_j, 2\vep),\\
    \text{\it smooth connection}, &\text {\it  elsewhere.}
\end{array}
\right.
  \eerr

  Then,
  \be
 \bar{ f}_\vep=\frac{1}{|\Omega|}\ito f_\vep\ud x\le \frac{4\pi N\vep^2}{|\Omega|}. \label{b.9}
  \ee
Define
\[g_\vep= \frac{8\pi N}{|\Omega|}(f_\vep-\bar{f}_\vep).\]
 It is easy to see that \[\ito g_\vep\ud x=0.\] Then we  see that the equation
 \be
  \Delta w=g_\vep \label{b.10}
 \ee
admits a unique solution up to an additive constant.

 First, it follows from \eqref{b.9} that, for $x\in B(p_j, \vep)$,
\be
 g_\vep\ge \frac{4\pi N}{|\Omega|}\left(2-\frac{8\pi N\vep^2}{|\Omega|}\right)> \frac{4\pi N}{|\Omega|}\label{b.11}
\ee if $\vep$ is small enough. In the sequel we fix $\vep$ such that
\eqref{b.11} is valid.

   Next, we choose a solution of \eqref{b.10}, say, $w_0$, to satisfy
   \[\re^{u_0+w_0}\le 1, \,\, x\in \Omega.\]
Hence, for any $\lambda>0,$ we have
 \be
 \Delta w_0=g_\vep> \frac{4\pi N}{|\Omega|}\ge \lambda\re^{u_0+w_0}(\re^{u_0+w_0}-1)^5+ \frac{4\pi N}{|\Omega|}\label{b.12}
 \ee
for $x\in B(p_j, \vep), \, j=1, 2, \cdots, N.$

 Finally, set
 \berr
  \mu_0=\inf\left\{\re^{u_0+w_0}\left| x\in \Omega\setminus\bigcup\limits_{j=1}^N B(p_j, \vep),\right.\right\}\\
  \mu_1=\sup\left\{\re^{u_0+w_0} \left| x\in \Omega\setminus\bigcup\limits_{j=1}^N B(p_j, \vep)\right.\right\}.
 \eerr
Then $0<\mu_0<\mu_1$ and $\re^{u_0+w_0}(\re^{u_0+w_0}-1)^5\le
\mu_0(\mu_1-1)^5=-C_0<0$ for
 $x\in \Omega\setminus\bigcup\limits_{j=1}^N B(p_j, \vep)$.
  As a consequence, we can choose $\lambda>0$ sufficiently large to fulfill \eqref{b.12} in entire $\Omega$.
  Thus, $w_0$ is a sub-solution of \eqref{b.3}. The proof of Lemma
  \ref{lem2} is complete.

Now we seek the critical value of the coupling parameter. We
establish the following lemma.

\begin{lemma}\label{lem3}
 There is a critical value of $\lambda$, say,  $\lambda_c$, satisfying
 \be
  \lambda_c\ge \frac{6^6}{5^5}\frac{4\pi N}{|\Omega|}, \label{b.13}
 \ee
  such that, for $\lambda>\lambda_c,$ the equation \eqref{b.3} has a solution, while for $\lambda<\lambda_c$,
  the equation  \eqref{b.3} has no solution.
\end{lemma}

{\it Proof.} \quad  Assume that $v$ is a solution of \eqref{b.3}.
Then $u=u_0+v$ satisfies \eqref{2.14'} and is negative near the
points $x=p_j, j=1, \cdots, N$. Applying  the maximum principle away
from the points $x=p_j, j=1, \cdots, N$, we see that $u<0$
throughout $\Omega$.

 Define
 \[ \Lambda=\big\{\lambda>0\big| \lambda \,\,\text{is such that \eqref{b.3} has a solution}\big\}.\]
Then we can prove  that  $\Lambda$ is an interval. To do so, we
prove that, if $\lambda'\in \Lambda$, then $[\lambda',
+\infty)\subset\Lambda$. Denote by $v'$ the solution of \eqref{b.3}
at $\lambda=\lambda'$. Noting that $u_0+v'<0$, we see that $v'$ is a
sub-solution of \eqref{b.3} for any $\lambda>\lambda'$. By Lemma
\ref{lem1},  we obtain $\lambda\in \Lambda$.

Let $\lambda_c=\inf\Lambda.$ Then, by the necessary condition
\eqref{b.5}, we have  $\lambda>\frac{6^6}{5^5}\frac{4\pi
N}{|\Omega|}$ for any $\lambda>\lambda_c$. Taking the limit
$\lambda\to\lambda_c$, we obtain \eqref{b.13}. Then Lemma \ref{lem3}
follows.

Now we need to  consider the critical case $\lambda=\lambda_c$. We
use the method of \cite{taran96} to deal with this  case.

We first make a simple observation.  We can show that the maximum
solutions of \eqref{b.3} $\{v_\lambda |\lambda>\lambda_c\}$ are
monotone family in the sense that $v_{\lambda_1}>v_{\lambda_2}$
whenever $\lambda_1>\lambda_2>\lambda_c$. Indeed, since
$u_0+v_\lambda<0$, from \eqref{b.3} we have
 \berr
 \Delta v_{\lambda_2}&=&\lambda_2\re^{u_0+v_{\lambda_2}}(\re^{u_0+v_{\lambda_2}}-1)^5+\frac{4\pi N}{|\Omega|}\\
&=&\lambda_1\re^{u_0+v_{\lambda_2}}(\re^{u_0+v_{\lambda_2}}-1)^5+\frac{4\pi N}{|\Omega|}+(\lambda_2-\lambda_1)\re^{u_0+v_{\lambda_2}}(\re^{u_0+v_{\lambda_2}}-1)^5\\
&\ge&\lambda_1\re^{u_0+v_{\lambda_2}}(\re^{u_0+v_{\lambda_2}}-1)^5+\frac{4\pi
N}{|\Omega|}
 \eerr
for $\lambda_1>\lambda_2>\lambda_c$. Therefore $ v_{\lambda_2}$ is a
sub-solution of \eqref{b.3}  with $\lambda=\lambda_1$.  Hence
$v_{\lambda_1}>v_{\lambda_2}$ if $\lambda_1>\lambda_2>\lambda_c$.

Set \[X=\left\{v\in W^{1,2}(\Omega)\Bigg| \ito v\ud x=0\right\}.\]
Then $X$ is a closed subspace of $W^{1,2}(\Omega)$ and
\[W^{1,2}(\Omega)=\mathbb{R}\oplus X.\]
In other words, for any $v\in W^{1,2}(\Omega)$, there exits a unique
number $c\in \mathbb{R}$ and $v'\in X$ such that $v=c+v'$.  In what
follows, we will use the  Trudinger--Moser  inequality(see
\cite{aubi})
 \be
   \ito\re^{v'}\ud x\le C\exp{\left(\frac{1}{16\pi}\ito|\nabla v'|^2\ud x\right)},\quad \forall\, v'\in X,\label{b13'}
 \ee
  where $C$  is a  positive constant depending only on $\Omega$.

\begin{lemma}\label{lem5}
 Let $v_\lambda$ be a solution of \eqref{b.3}. Then $v_\lambda=c_\lambda+v'_\lambda$,
 where  $c_\lambda\in \mathbb{R}$ and $v'_\lambda\in X$. We have
 \be
 \|\nabla v'_\lambda\|_2\le C\lambda,\label{b.14}
 \ee
 where $C$ is a positive constant depending only on the size of the
 torus $\Omega$. Furthermore, $\{c_\lambda\}$ satisfies the estimate
 \be|c_\lambda|\le C(1+\lambda+\lambda^2).\label{b.15}\ee
 Especially, $v_\lambda$ satisfies
 \be\|v_\lambda\|_{W^{1,2}(\Omega)}\le C(1+\lambda+\lambda^2).\label{b.16}\ee
\end{lemma}

{\it  Proof.}\quad  Multiplying \eqref{b.3} by $v'_\lambda$,
integrating over $\Omega$, using Schwarz  inequality and
Poincar\'{e} inequality, we have
 \berr
 \|\nabla v'_\lambda\|_2^2&=&-\ito \lambda\re^{u_0+v_{\lambda}}(\re^{u_0+v_{\lambda}}-1)^5 v'_\lambda\ud x\\
 &\le&2\lambda\ito|v'_\lambda|\ud x\le
 2\lambda|\Omega|^{1/2}\|v'_\lambda\|_2\le C\lambda\|\nabla v'_\lambda\|_2,
 \eerr
which implies \eqref{b.14}.

Noting the property $u_0+v_\lambda=u_0+c_\lambda+v'_\lambda<0$, we
have the upper bound,
 \be c_\lambda<-\frac{1}{|\Omega|}\ito u_0(x)\ud x.\label{b.17}\ee

From the equation \eqref{b.3}, we have
\[\Delta v_\lambda=\lambda\re^{u_0+v_{\lambda}}(\re^{u_0+v_{\lambda}}-1)^5+\frac{4\pi N}{|\Omega|}
\ge\lambda\re^{u_0+v_{\lambda}}(\re^{u_0+v_{\lambda}}-1)+\frac{4\pi
N}{|\Omega|}.\] Integrating  the above inequality over $\Omega$
gives
\[\lambda\ito \re^{u_0+v_{\lambda}}\ud x\ge \lambda \ito \re^{2(u_0+v_{\lambda})}\ud x+4\pi N>4\pi N,\]
which is
\[\lambda \re^{c_\lambda}\ito\re^{u_0}\re^{v'_\lambda}\ud x\ge 4\pi N.\]
Then we have

\berr
 \re^{c_\lambda}&\ge& C\lambda^{-1}\left(\ito\re^{u_0}\re^{v'_\lambda}\ud x\right)^{-1}
 \ge C\lambda^{-1}\left(\ito\re^{v'_\lambda}\ud x\right)^{-1}
  \ge C\lambda^{-1}\exp\left(-\frac{1}{16\pi}\|\nabla v'_\lambda\|_2^2\right),
\eerr
 where in the last inequality  we have used Trudinger--Moser \eqref{b13'}.

  Now using \eqref{b.3} in the above inequality we can
 obtain a lower bound of $c_\lambda$,
  \be
  c_\lambda\ge -C(1+\lambda+\lambda^2).\label{b.18}
  \ee

  Then \eqref{b.15} follows from  \eqref{b.17} and \eqref{b.18}.
  Combining \eqref{b.14} and \eqref{b.15}, we obtain \eqref{b.16}.
  Hence Lemma \ref{lem5} follows.

 For the critical case we have the following result.
\begin{lemma}\label{thc1}
The set of $\lambda$ for which the equation \eqref{b.3} has a
solution is a closed interval. That is to say,  at
$\lambda=\lambda_c$ \eqref{b.3} has a solution as well.
\end{lemma}

{\it Proof.} \quad  For $\lambda_c<\lambda<\lambda_c+1$ (say), by
Lemma \ref{lem5} the set $\{v_\lambda\}$ is bounded in
$W^{1,2}(\Omega)$. Noting that $\{v_\lambda\}$ is monotone with
respect to $\lambda$, we conclude that there exist $v_\ast\in
W^{1,2}(\Omega)$ such that
 \[v_\lambda\to v_\ast    \quad \text{weakly in }\quad W^{1,2}(\Omega)\quad \text{as} \quad \lambda\to\lambda_c.\]
Therefore $v_\lambda\to v_\ast$ strongly in $L^p(\Omega)$ for any
$p\ge1$ as $\lambda\to\lambda_c$. Using Trudinger--Moser inequality
\eqref{b13'}  again we obtain $\re^{v_\lambda}\to \re^{v_\ast}$
strongly in $L^p(\Omega)$ for any $p\ge1$ as $\lambda\to\lambda_c$.
Using this result in \eqref{b.3} and the $L^2$ estimates for the
elliptic equations, we have $v_\ast\in W^{2,2}(\Omega)$ and
$v_\lambda\to v_\ast$ strongly in $W^{2, 2}(\Omega)$ as
$\lambda\to\lambda_c$. Particularly, taking the limit
$\lambda\to\lambda_c$ in \eqref{b.3}, we obtain that $v_\ast$ is a
solution of \eqref{b.3} for $\lambda=\lambda_c$. Then the lemma
follows.

Denote
\[P=\{p_1, \cdots, p_m; n_1, n_2, \cdots, n_m\},\qquad P'=\{p_1, \cdots, p_m; n'_1, n'_2, \cdots, n'_m\}.\]
 We denote the dependence of $\lambda_c$ on $P$ by $\lambda_c(P)$.
 Consider the equation
 \be
  \Delta u=\lambda\re^u(\re^u-1)^5+4\pi\sum\limits_{j=1}^mn_j\delta_{p_j}\label{b.19}
 \ee

\begin{lemma}\label{lem4}
 If $P\le P'$, we have $\lambda(P)\le \lambda(P')$.
\end{lemma}

 {\it Proof.}\quad It is sufficient to show that, if $\lambda>\lambda_c(P')$, then $\lambda\ge\lambda_c(P)$.
 Let $u'$ be a solution of \eqref{b.19} with $n_j=n'_j, \, j=1, \cdots, m$ and $u_0$ satisfy
 \[\Delta u_0=-\frac{4\pi N}{|\Omega|}+4\pi\sum\limits_{j=1}^mn_j\delta_{p_j},\]
where $N=n_1+\cdots+n_m$. Setting $u'=u_0+v\_$,  we have

\[ \Delta v\_=\lambda\re^{u_0+v\_}(\re^{u_0+v\_}-1)^5+\frac{4\pi N}{|\Omega|}+4\pi\sum\limits_{j=1}^m(n'_j-n_j)\delta_{p_j},\]
which implies in particular that $v\_$ is a sub-solution of
\eqref{b.3} in the sense of distribution and \eqref{b.7} holds
pointwise. It is easy to check that the singularity of $v\_$ is at
most of the type $\ln|x-p_j|$. Hence, the inequality \eqref{b.7}
still results in the convergence of the sequence of $\{v_n\}$ to a
solution of \eqref{b.3} in any $C^k$ norm. Indeed, by \eqref{b.7} we
see that $\{v_n\}$ converges almost everywhere and is bounded in
$L^2$ norm. Therefore, the sequence converges in $L^2$.
 Analogously, the right-hand side of \eqref{b.6} also converges in $L^2$. Applying the standard $L^2$ estimate,
  we see that the sequence converges in $W^{2,2}(\Omega)$ to a strong solution of \eqref{b.3}.
  Thus, a classical solution can be obtained. Using a bootstrap argument again, we can obtain the convergence in $C^k$ norm.
   This proves $\lambda\ge \lambda_c(P)$. Therefore , $\lambda(P)\le \lambda(P')$.

 From the above discussion we  complete the proof of Theorem
 \ref{th1}.

  Now we carry out the proof of Theorem \ref{th2}.
  It is easy to see that \eqref{b.3} is the Euler--Lagrangian
  equation of the following functional
   \be
   I_\lambda(v)=\int_{\Omega}\left\{\frac12|\nabla v|^2+\frac{\lambda}{6}(e^{u_0+v}-1)^6+\frac{4\pi N}{|\Omega|}v\right\}\ud x.\label{b.20}
   \ee

\begin{lemma}\label{l1.6}
  For every $\lambda>\lambda_c$, the problem \eqref{b.3} admits a
  solution $v_\lambda\in W^{1,2}(\Omega)$ and it is a local minimum of the functional
  $I_\lambda(v)$
  defined by \eqref{b.20}.
\end{lemma}

{\it Proof.}\quad  We  apply the method in  \cite{taran96}. Since
$u_0+v_\ast<0$, we see
  that $v_\ast$ is a sub-solution of \eqref{b.3} for any
  $\lambda>\lambda_c$.  Define
  \be
  V=\big\{v\in W^{1, 2}| v\ge v_\ast  \quad \text{a. e.  in }\,\, \Omega\big\}. \label{b.21}
  \ee
  Then the functional $I_\lambda$ is bounded form below on $V$.
 We can study the following minimization problem
  \be
  \eta_0\equiv\inf\big\{I_\lambda(v)|v\in V\big\}.\label{b.22}
  \ee
 We will show that the problem \eqref{b.22} admits a solution.

 Let $\{v_n\}$ be a minimizing sequence of \eqref{b.22}. Then, by
 the decomposition formula,
  \[
  v_n=v_n'+c_n, \quad  v_n'\in X,\,\,\, c_n\in \mathbb{R}, \quad  n=1, 2,
  \dots,
   \]
 we see that $\{\|\nabla v_n\|_2\}$ is bounded since the definition
 of $V$ gives a lower bound of $\{c_n\}$. By  the definition of $I_\lambda(v)$
 we have
 \[ I_\lambda(v_n)\ge 4\pi c_n,\]
which gives an upper bound of $\{c_n\}$. Then $\{v_n\}$ is a bounded
sequence in $W^{1,2}(\Omega)$. Without loss of generality, we may
assume that $\{v_n\}$ converges weakly to an element  $v\in
W^{1,2}(\Omega)$ as $n\to\infty$. Hence, $v$ is a solution to the
problem \eqref{b.22}. Using Lemma 5.6.3 in \cite{yang1} or the
appendix of  \cite{taran96}, we conclude that $v$ is a solution of
the equation \eqref{b.3} and $v\ge v_\ast$ in $\Omega$. By the
maximum principle we obtain the strict inequality $v>v_\ast$ in
$\Omega$.

 Next we prove that $v$ is a local minimum of the functional
 \eqref{b.20} in $W^{1, 2}(\Omega)$.  We use the approach of Brezis
 and Nirenberg \cite{brni} as in Tarantello \cite{taran96} and Yang \cite{yang1}. We argue by
 contradiction. Suppose otherwise $v$ is not a local minimum of
 $I_\lambda(v)$ in $W^{1,2}(\Omega)$.   Then, for any integer
 $n\ge1$, we have
 \be
  \inf\left\{I_\lambda(w)| w\in W^{1, 2}(\Omega), \|w-v\|_{W^{1, 2}(\Omega)}\le
  \frac1n\right\}=\vep_n<I_\lambda(v).\label{b.23}
  \ee
Similarly to the  above, for any $n\ge1$,  we can conclude that the
infimum of \eqref{b.23} is achieved at a point $v_n \in
W^{1,2}(\Omega)$. Then, by the principle of Larangian multipliers,
we obtain that there exists  number $\mu_n\le0$ such that
 \[ -\Delta v_n+\lambda e^{u_0+v_n}(e^{u_0+v_n}-1)^5+\frac{4\pi N}{|\Omega|}=\mu_n(-\Delta[v_n-v]+v_n-v).\]
 We rewrite the above equation the following form
 \ber
 \Delta(v_n-v)=\frac{\lambda}{1+|\mu_n|}\left[e^{u_0+v_n}(e^{u_0+v_n}-1)^5-e^{u_0+v}(e^{u_0+v}-1)^5\right]+\frac{|\mu_n|}{1+|\mu_n|}(v_n-v).\label{b.24}
 \eer
Noting the fact $\|v_n-v\|_{W^{1,2}(\Omega)}\to 0$ as $n\to \infty$
and the  Trudinger--Moser inequality \eqref{b13'}, we see that the
right hand side of \eqref{b.24} converges to $0$ as $n\to \infty$.
Then using the elliptic $L^2$ estimate, we have $v_n\to v$ in
$W^{2,2}(\Omega)$ as $n\to \infty$. By embedding theorem we see that
$v_n\to v$ in $C^\alpha(\Omega)$ for any $0<\alpha<1$. Since
$\Omega$ is compact and $v>v_\ast$ in $\Omega$, we have $v_n>v_\ast$
for $n$ sufficiently large. This implies $v\in V$ for  $n$
sufficiently large, which leads to $I_\lambda(v_n)\ge I_\lambda(v)$.
Then we obtain a contradiction and the conclusion follows.

In the sequel we show that the functional $I_\lambda(v)$ satisfies
P.S. condition in $W^{1,2}(\Omega)$.

\begin{lemma}\label{l1.7}
 Any sequence $\{v_n\} \subset W^{1,2}(\Omega) $ verifying
 \be
 I_\lambda(v_n)\to \alpha, \quad   \|dI_\lambda(v_n)\|_d\to 0     \quad as \,\, n\to
 \infty\label{b.25}
 \ee
 admits a convergent subsequence, where we use $\|\cdot\|_d$ to denote
 the norm of the dual space of $W^{1,2}(\Omega)$.
\end{lemma}

{\it Proof.}\quad
   By \eqref{b.25} we have
   \ber
   &&\frac12\|\nabla v_n\|_2^2+\frac{\lambda}{6}\ito(e^{u_0+v_n}-1)^6\ud x+\frac{4\pi N}{|\Omega|}\ito   v_n\ud x\to \alpha, \label{b.26}\\
  &&\left|\ito\nabla v_n\nabla\varphi\ud x+\lambda\ito e^{u_0+v_n}(e^{u_0+v_n}-1)^5\varphi\ud x+\frac{4\pi N}{|\Omega|}\ito\varphi\ud x\right|
  \le
  \vep_n\|\varphi\|_{W^{1,2}(\Omega)} \label{b.27}
   \eer
  as $n\to \infty$,  where $\vep_n\to 0$ as $n\to \infty$.
 Setting $\varphi=1$ in \eqref{b.27}, we obtain
 \[ \lambda\ito e^{u_0+v_n}(e^{u_0+v_n}-1)^5\ud x +4\pi N\le \vep_n|\Omega|,\]
which implies
 \berr
 \frac{\vep_n|\Omega|}{\lambda}&\ge&\frac{4\pi N}{\lambda}+\ito e^{u_0+v_n}(e^{u_0+v_n}-1)^5\ud x\\
 &=&\frac{4\pi N}{\lambda}+\ito (e^{u_0+v_n}-1)^6\ud x+\ito (e^{u_0+v_n}-1)^5\ud x\\
 &\ge&\frac{4\pi N}{\lambda}-\frac56|\Omega|+\frac56\ito (e^{u_0+v_n}-1)^6\ud x.
 \eerr
Then, it follows
 \be
 \ito (e^{u_0+v_n}-1)^6\ud x\le
 \left(1+\frac65\frac{\vep_n}{\lambda}\right)|\Omega|-\frac{24\pi
 N}{5}\le C.\label{b.28}
 \ee
 Here  and in the sequel we use $C$ to denote a universal positive constant maybe different in different places.
Hence, by \eqref{b.28} we have
 \be
 \ito e^{6(u_0+v_n)}\ud x=\ito [(e^{u_0+v_n}-1)+1]^6\ud x\le
 2^5\left[\ito(e^{u_0+v_n}-1)^6\ud x+|\Omega|\right]\le C.\label{b.29}
 \ee
Using H\"{o}lder inequality and \eqref{b.29}, we have
 \be
 \ito e^{2(u_0+v_n)}\ud x\le\left(\ito e^{6(u_0+v_n)}\ud
 x\right)^{\frac13}|\Omega|^{\frac23}\le C.\label{b.30}
 \ee
 Likewise,
  \be
 \ito e^{4(u_0+v_n)}\ud x\le C.\label{b.31}
 \ee

Applying the decomposition formula $v_n=v_n'+c_n$ in \eqref{b.26},
we have
 \be
  \frac12\|\nabla v_n'\|_2^2+\frac{\lambda}{6}\ito(e^{u_0+v_n'+c_n}-1)^6\ud x+4\pi Nc_n\to \alpha, \label{b.32}
 \ee
as $n\to \infty$.
 Then from \eqref{b.32} it follows that $c_n$ is bounded from above.
Since $I_\lambda(v_n)\to \alpha$ as $n\to \infty$, we may assume
that for all $n$,
\[\alpha-1<I_\lambda(v_n)<\alpha+1,\]
which leads to
 \be
  \alpha-1<\frac12\|\nabla v_n'\|_2^2+\frac{\lambda}{6}\ito(e^{u_0+v_n'+c_n}-1)^6\ud x+4\pi Nc_n<\alpha+1.\label{b.33}
 \ee
Therefore it follows from \eqref{b.29} and \eqref{b.33} that
 \be
  \alpha-1+\frac{4\lambda\pi N}{5}-\left(\frac\lambda6+\frac{\vep_n}{5}\right)|\Omega|<\frac12\|\nabla v_n'\|_2^2+4\pi Nc_n<\alpha+1.\label{b.34}
 \ee

Now we aim to get a lower bound for $c_n$. Let $\varphi=v_n'$
in\eqref{b.27}, we obtain
  \[
   \|\nabla v_n'\|_2^2+\lambda\ito  e^{u_0+v_n}(e^{u_0+v_n}-1)^5v_n'\ud  x\le \vep_n\|v_n'\|_{W^{1,2}(\Omega)}\le C\vep_n\|\nabla v_n'\|_2
  \]
from which follows
 \ber
   &&\|\nabla v_n'\|_2^2+\lambda\ito  e^{6(u_0+v_n)}v_n'\ud  x\nonumber\\
   &&\le  C\ito e^{u_0+v_n}(e^{4(u_0+v_n)}+e^{3(u_0+v_n)}+e^{2(u_0+v_n)}+e^{u_0+v_n}+1)|v_n'|\ud x+C\vep_n\|\nabla v_n'\|_2.\label{b.35}
  \eer
It is easy to see  that \eqref{b.35} is equivalent to
  \ber
   &&\|\nabla v_n'\|_2^2+\lambda\ito  e^{6(u_0+c_n)}(e^{6v_n'}-1)v_n'\ud  x\nonumber\\
   &&\le \lambda\ito e^{6(u_0+c_n)}v_n'\ud x+C\vep_n\|\nabla v_n'\|_2\nonumber\\
  &&\quad + C\ito e^{u_0+v_n}(e^{4(u_0+v_n)}+e^{3(u_0+v_n)}+e^{2(u_0+v_n)}+e^{u_0+v_n}+1)|v_n'|\ud x.\label{b.36}
  \eer
 Now we deal the right hand side terms in \eqref{b.36}.
 Using the  H\"{o}ler inequality and the Poincar\'{e} inequality, we have
 \[
  \ito e^{6(u_0+c_n)}v_n'\ud x\le C\|v_n'\|_2\le C\|\nabla v_n'\|_2.
  \]
Applying H\"{o}ler inequality, \eqref{b.29} and Sobolev embedding
theorem, we get
 \[
 \ito(e^{5(u_0+v_n)}|v_n'|\ud x\le\left(\ito e^{6(u_0+v_n)}\ud x\right)^{\frac56}\left(\ito |v_n'|^6\ud x\right)^{\frac16}
 \le C\|v_n'\|_6\le C\|\nabla v_n'\|_2
 \]
 All the other terms on the right hand side of \eqref{b.36} can
 estimated in the same way and they all be bounded by $C\|\nabla v_n'\|_2$. Then we have
  \be
   \|\nabla v_n'\|_2^2+\lambda\ito  e^{6(u_0+c_n)}(e^{6v_n'-1})v_n'\ud x\le C\|\nabla v_n'\|_2.\label{b.37}
   \ee
 Noting that   \[\ito  e^{6(u_0+c_n)}(e^{6v_n'-1})v_n'\ud  x\ge 0,\]
 we obtain from \eqref{b.37} that
   \be
    \|\nabla v_n'\|_2\le C . \label{b.38}
   \ee
 Inserting \eqref{b.38} into \eqref{b.34}, we see that $c_n$ is
 bounded from below.

 Then we can derive that $\{v_n\}$ is uniformly bounded in
 $W^{1,2}(\Omega)$. Without loss of generality, we may assume that
 there exists an element $v\in W^{1,2}(\Omega)$ such that
  $v_n\to v$ weakly in $W^{1,2}(\Omega)$ and strongly in
  $L^p(\Omega)$ for any $p\ge 1$.

 Setting $n\to\infty$ in \eqref{b.27}, we have
 \be
 \ito \left\{\nabla v\cdot\nabla \varphi+\lambda e^{u_0+v}(e^{u_0+v}-1)^5\varphi+\frac{4\pi N}{|\Omega|}\varphi\right\}\ud x=0,
 \,\forall\, \varphi\in W^{1, 2}(\Omega). \label{b.39}
 \ee
Then $v$ is a critical point of the functional $I_\lambda$.

Next we show that $v_n\to v$ strongly in $W^{1,2}(\Omega)$ as
$n\to\infty$.

 Letting $\varphi=v_n-v$ in \eqref{b.27} and \eqref{b.39} and
 subtract the resulting expressions, we obtain
  \berr
  &&\|\nabla(v_n-v)\|_2^2+\lambda\ito\left[e^{u_0+v_n}(e^{u_0+v_n}-1)^5-e^{u_0+v}(e^{u_0+v}-1)^5\right](v_n-v)\ud  x\\
  && \le\vep_n\|v_n-v\|_{W^{1, 2}(\Omega)},
  \eerr
which implies
   \ber
  &&\|\nabla(v_n-v)\|_2^2\nonumber\\
  && \le\lambda\ito\left|e^{u_0+v_n}(e^{u_0+v_n}-1)^5-e^{u_0+v}(e^{u_0+v}-1)^5\right||v_n-v|\ud x+\vep_n\|v_n-v\|_{W^{1, 2}(\Omega)}\nonumber\\
  &&\le C\|v_n-v\|_2+\vep_n(\|v_n\|_{W^{1, 2}(\Omega)}+\|v\|_{W^{1, 2}(\Omega)}). \label{b.40}
  \eer
Since the right hand side of \eqref{b.40} tends to $0$ as $n\to
\infty$, we have  $\nabla v_n\to \nabla v$ strongly in
$L^2(\Omega)$. Then we can obtain that $v_n\to v$ strongly in
$W^{1,2}(\Omega)$ as $n\to \infty$. Then the proof of Lemma
\ref{l1.7} is complete.

Next we establish the existence of secondary solutions of the
equation \eqref{b.3}.

Let $v_\lambda$ be the  local minimum of $I_\lambda$ obtained in
Lemma \ref{l1.6}. Then There exists a positive constant $\delta >0$
such that
 \[
  I_\lambda(v_\lambda)<I_\lambda(w), \quad \forall\,  w\in W^{1,2}(\Omega),\quad \|w-v_\lambda\|_{W^{1,2}(\Omega)}\le \delta.
 \]
 Here we assume that $v_\lambda$ is a strict local minimum because
 otherwise we would already have additional solutions. Therefore we
 can assume that there admits a positive constant $\delta_0>0$ such
 that
  \[
  \inf\left\{I_\lambda(w)|  w\in W^{1, 2}(\Omega), \quad \|w-v_\lambda\|_{W^{1, 2}(\Omega)}=\delta_0\right\}>I_\lambda(v_\lambda).
  \]

We will show that the functional $I_\lambda$ possesses a ``mountain
pass" structure. Indeed, since $u_0+v_\lambda<0$, we have
 \berr
 I_\lambda(v_\lambda-c)-I_\lambda(v_\lambda)&=&\frac\lambda6\ito\left[(e^{u_0+v_\lambda-c}-1)^6-(e^{u_0+v_\lambda}-1)^6\right]\ud x
 -4\pi Nc\\
 &<& \frac\lambda6|\Omega|-4\pi Nc\to -\infty\quad
 \text{as}\quad c\to +\infty.
 \eerr
 Then we can choose $c_0>\delta_0$ sufficiently large such that
 \[
 I_\lambda(v_\lambda-c_0)<I_\lambda(v_\lambda)-1\quad
 \text{and}\quad |\Omega|^{\frac12}c_0>\delta_0.
 \]

 Denote by $\mathcal {P}$ the set of all continuous paths in
 $W^{1,2}(\Omega)$  $\gamma(t): [0, 1]\rightarrow W^{1,2}(\Omega)$
 connecting the points $v_\lambda$ and $v_\lambda-c_0$ with $\gamma(0)=v_\lambda, \,
 \gamma(1)=v_\lambda-c_0$.  Define
  \[
  \alpha_0=\inf\limits_{\gamma\in\mathcal{P}}\sup\limits_{t\in[0,1]}\big\{I_\lambda(\gamma(t))\big\}.
  \]
Then we have
 \be\alpha_0>I_\lambda(v_\lambda).\label{b.41}\ee
Therefore the functional  $I_\lambda$ satisfies all the hypothesis
of the mountain pass theorem of Ambrosetti-Rabinowitz \cite{amra}.
Then we can conclude that $\alpha_0$ is a critical value of the
functional $I_\lambda$  in $W^{1,2}(\Omega)$. Noting \eqref{b.41},
we have an additional solution of the equation \eqref{b.3}. Then we
 complete the proof of  Theorem \ref{th2}.

\section{ Existence of topological solutions}\label{s5}

\setcounter{equation}{0}
 \setcounter{lemma}{0}

In this section we establish the existence of topological solution
of  the generalized self-dual Chern--Simons equations
\eqref{2.8}-\eqref{2.9}, i.e. we prove Theorem \ref{th3}. We will
use a super- and sub-solution method to construct solutions. The key
step is to find a suitable  sub-solution to the reduced equation.
This technique maybe applied to the problems with similar
structures.

As in Section \ref{s3}, let $|\phi|^2=\re^u$,   the prescribed zeros
of $\phi$ be $p_1, \dots, p_m$ with multiplicities $n_1, \dots,
n_m$, respectively, and $N=\sum_{s=1}^mn_s$.  Then  we arrive at the
following governing equation
   \be
    \Delta  u=\lambda\mathrm{e}^u(\mathrm{e}^u-1)^5+4\pi\sum\limits_{s=1}^mn_s\delta_{p_s},  \quad x\in\mathbb{R}^2.\label{c.1'}
   \ee
Correspondingly, the topological condition \eqref{2.10}  is  changed
 into
   \be
   u(x)\to 0\quad\text{as}\quad |x|\to +\infty.\label{c.2'}
  \ee

We define the background function
 \be
 u_0(x)=-\sum\limits_{s=1}^mn_s\ln(1+|x-p_s|^{-2}). \label{c.1}
 \ee
Then
 \be
 \Delta
 u_0=4\pi\sum\limits_{s=1}^{m}n_s\delta_{p_s}-g,\label{c.2}
 \ee
where $g=4\sum\limits_{s=1}^{m}n_s(1+|x-p_s|^2)^{-2}$.
 Let $v=u-u_0$, we have
 \be
 \Delta v=\lambda \re^{u_0+v}(\re^{u_0+v}-1)^5+g.\label{c.3}
 \ee
 The boundary condition \eqref{c.2'} becomes
  \be
  v(x)\to0 \qquad \text{as}\quad |x|\to +\infty.\label{c.4}
  \ee

It is easy to check that $v^\ast=-u_0$ is an super-solution to the
problem \eqref{c.3}-\eqref{c.4}.

Next we construct a sub-solution to the
problem\eqref{c.3}-\eqref{c.4}. The construction of sub-solution is
a crucial part of the proof. This technique maybe applied to the
other problems with similar structures.

\begin{lemma}\label{lemc2}
 For any $\lambda>0$, the problem \eqref{c.3}-\eqref{c.4} admit a
 sub-solution.
\end{lemma}

 {\it  Proof. }\quad It is shown in
\cite{spya4,wangr} that
 for any $\mu>0$ the equation
  \be
  \Delta u=\mu \mathrm{e}^u(\mathrm{e}^u-1)+4\pi\sum\limits_{s=1}^mn_s\delta_{p_s}\label{c.5}
  \ee
has a topological solution $u^{(\mu)}$, namely,  $u^{(\mu)}$
satisfies the equation \eqref{c.5}  and the boundary condition
$u^{(\mu)}\to 0$ as $|x|\to+\infty$. Then for $\mu= \lambda
\mathrm{e}^{-a}(\mathrm{e}^{-a}-1)^4, \quad a>0$, the problem
\eqref{c.5} admits a solution $u_\ast$  satisfies
  \be
  \Delta u_\ast =\lambda(\mathrm{e}^{-a}-1)^4\mathrm{e}^{u_\ast-a}(e^{u_\ast}-1)+4\pi\sum\limits_{s=1}^mn_s\delta_{p_s}\label{c.6}
  \ee
   and $u_\ast\to 0$ as $|x|\to +\infty$ and $u_\ast<0$ for all $x\in  \mathbb{R}^2$.
Since $u_\ast<0,$  for any $a>0$, we have
  \[\mathrm{e}^{u_\ast-a}-1<\mathrm{e}^{-a}-1<0\]
and
  \[\mathrm{e}^{u_\ast}-1>\mathrm{e}^{u_\ast-a}-1\]
 Then we obtain from \eqref{c.6} that
   \be
  \Delta (u_\ast-a)\ge\lambda  \mathrm{e}^{u_\ast-a}(\mathrm{e}^{u_\ast-a}-1)^5+4\pi\sum\limits_{s=1}^mn_s\delta_{p_s} \label{c.7}
  \ee
Let $v_\ast=u_\ast-a-u_0,$  from \eqref{c.7} we have
  \be
  \Delta v_\ast\ge\lambda  \mathrm{e}^{u_0+v_\ast}(\mathrm{e}^{u_0+v_\ast}-1)^5+g \label{c.8}
  \ee
  and  $v_\ast$ satisfies $v_\ast\to -a$ as $|x|\to \infty$.
Then we conclude that $v_\ast$ is a sub-solution to the problem
\eqref{c.3}-\eqref{c.4}. Then the lemma follows.

At this point we can establish a solution of  to the problem
\eqref{c.3}-\eqref{c.4} by the super-solution $v^*$ and sub-solution
$v_*$.

 Let $B_r$ be  a ball centered at the origin with radius $r$ in $\mathbb{R}^2$, where $r>|p_s|, s=1,\dots, m$.
  Consider the following boundary value problem
   \ber
    \Delta v=\lambda \mathrm{e}^{u_0+v}(\mathrm{e}^{u_0+v}-1)^5+g, \quad x\in B_r,\label{c.9}\\
    v=-u_0, \quad x\in \partial B_r. \label{c.10}
   \eer
We first prove that the problem \eqref{c.9}-\eqref{c.10} has a
unique solution $v$ satisfying $v_*<v<v^*$.

It is easy to see that $v^\ast=-u_0$ and $v_\ast=u_\ast-a-u_0$ are a
pair of ordered  super-and sub-solutions to the problem
\eqref{c.9}-\eqref{c.10}.

We use the monotone iterative method.
    Let $K>0$ be constant satisfying $K\ge 6\lambda$. We
 first introduce an iteration sequence on $B_r$.
  \ber
  (\Delta-K)v_n&=&\lambda \re^{u_0+v_{n-1}}(\re^{u_0+v_{n-1}}-1)^5-Kv_{n-1}+g \quad  \text{in }\quad B_r,\label{c.11}\\
  v_n&=&v^* \quad\text{on }\quad\partial B_r,\,\, n=1,2,\cdots,\label{c.12}\\[1mm]
  v_0&=&v^*.\label{c.13}
  \eer
\begin{lemma}\label{lemc1}
Let $\{v_n\}$ be the sequence defined by the iteration scheme
\eqref{c.11}. Then
 \be
 v^*> v_1> v_2> \cdots> v_n> \cdots >v_*. \label{c.14}
\ee
\end{lemma}

{\it Proof.}\quad   We prove this lemma by induction.

   For $n=1$, $v_1$ satisfies
   \be
   (\Delta-K)v_1=-Kv^*+g \quad  \text{in }\quad B_r.\label{c.15'}
  \ee
 It is easy to see that the right hand side of \eqref{c.12} belongs
to $L^p(B_r)$ for $p>2$. Then by the standard theory, we have
$v_1\in C^{1, \alpha}(\overline{B_r}) (0<\alpha<1)$. Near the set
$Q=\{p_1,\dots,p_m\}$ we have $v_1<v^*$. In $B_r\backslash Q$, we
can get
 \[\Delta (v_1-v^*)-K(v_1-v^*)=0.\]
Then, by maximum principle we have $v_1<v^*$ in $B_r$.
 Noting that $v_*<v^*$, we have
\berr(\Delta-K)(v_*-v_1)&\ge&\lambda\mathrm{e}^{u_0+v_*}(\mathrm{e}^{u_0+v_*}-1)^5-K(v_*-v_1)\\
&=&
[\lambda\mathrm{e}^{u_0+\xi}(\mathrm{e}^{u_0+\xi}-1)^4(6\mathrm{e}^{u_0+\xi}-1)-K](v_*-v_1)\\
&\ge& (5\lambda-K)(v_*-v_1)\\
&\ge&0.
 \eerr
Here and what after we use $\xi$ to denote an intermediate quantity
from the mean value theorem. Hence by maximum principle again we
have $v_*<v_1$ in $B_r$.

Suppose that we have already obtained the inequality
 $v_*<v_k$,  $v_k<v_{k-1}$ for some $k\ge 1$. Then  by \eqref{c.11}
we have
 \berr
 (\Delta-K)(v_{k+1}-v_k)
&=&\lambda[\mathrm{e}^{u_0+v_k}(\mathrm{e}^{u_0+v_k}-1)^5-\mathrm{e}^{u_0+v_{k-1}}(\mathrm{e}^{u_0+v_{k-1}}-1)^5]-K(v_k-v_{k-1})\\
&=&[\lambda\mathrm{e}^{u_0+\xi}(\mathrm{e}^{u_0+\xi}-1)^4(6\mathrm{e}^{u_0+\xi}-1)-K](v_k-v_{k-1})\\
&\ge& (5\lambda-K)(v_k-v_{k-1})\\
&\ge&0.
 \eerr
Therefore we have $v_{k+1}<v_k$  in $B_r$ by maximum principle.
Similarly, we have
 \berr
 (\Delta-K)(v_*-v_{k+1})
&\ge&\lambda[\mathrm{e}^{u_0+v_*}(\mathrm{e}^{u_0+v_*}-1)^5-\mathrm{e}^{u_0+v_k}(\mathrm{e}^{u_0+v_k}-1)^5]-K(v_*-v_k)\\
&=&[\lambda\mathrm{e}^{u_0+\xi}(\mathrm{e}^{u_0+\xi}-1)^4(6\mathrm{e}^{u_0+\xi}-1)-K](v_*-v_k)\\
&\ge& (5\lambda-K)(v_*-v_k)\\
&\ge&0.
 \eerr
Hence we obtain $v_*<v_{k+1}$ in $B_r$.  Then, we get \eqref{c.14}.
Hence Lemma \ref{lemc1} follows.

Since $v_*$ is a bounded function, we can get the existence of  the
pointwise limit
 \be
 v=\lim\limits_{n\to\infty}v_n.\label{c.15}
 \ee
 Let $n\to \infty$ in \eqref{c.11} and by the elliptic estimate and
 embedding theorem we see that the limit \eqref{c.15} can be
achieved in any strong sense and $v$ is a smooth solution of
\eqref{c.9}-\eqref{c.10}. It is easy to see that the solution $v$ is
unique and $v$ satisfies $v_*<v<v^*$.

 Now we denote by $v^{(n)}$ the  solution of \eqref{c.9}-\eqref{c.10}
with $r=n (n$ is large such that $n>|p_s|, s=1,\dots, m)$. By the
construction of $v^{(n)}$, we have $v^{(n+1)}\le v^*$ in $\partial
B_{n+1}$.  Then, $v^{(n+1)}$ is a sub-solution of
\eqref{c.9}-\eqref{c.10} with $r=n$. Therefore, from Lemma
\ref{lemc1} we have $v^{(n+1)}\le v^{(n)}$ in $B_n$ for any $n$.
Then for each fixed $n_0\ge1$, we have the monotone sequence
$v_{n_0}>v_{n_0+1}>\dots>v_n>v_{n+1}>\dots>v_*$ in $B_{n_0}$. Then
we can see that the sequence $\{v^{(n)}\}$ converges to a solution,
say $v$, of the equation \eqref{c.3} over the full plane
$\mathbb{R}^2$. By elliptic  $L^p$ estimate, we have $v\in
W^{2,2}(\mathbb{R}^2)$. Then we get $v(x)\to 0$ as $|x|\to \infty$,
which is  the topological boundary condition \eqref{c.4}. Then we
can get a topological solution $u$ of \eqref{c.1'}  satisfying $u<0$
in $\mathbb{R}^2$.

 Now we show that $v$ is maximal.  Let $\tilde{v}$ be another
solution to \eqref{c.9}-\eqref{c.10}.  Then $\tilde{v}$ satisfies
 \[
 \Delta(u_0+\tilde{v})=\lambda\re^{u_0+\tilde{v}}(\re^{u_0+\tilde{v}}-1)^5
 \quad \text{in }\quad \mathbb{R}^2\setminus\{p_1,\dots, p_m\},
\]
 $u_0+\tilde{v}=0$ at infinity, and $u_0+v<0$ in a small
 neighborhood of $\{p_1,\dots, p_m\}$. Using maximum principle, we
 see that $u_0+\tilde{v}\le 0$. Then by Lemma \ref{lemc1}, we obtain
 $\tilde{v}\le v$, which is to  say that   $v$ is maximal.

Let $u$ be the solution of \eqref{c.1'} obtained above.  Define
  \berr
  &&\theta(z)=\sum\limits_{s=1}^mn_s\arg(z-z_s), \quad \phi(z)=\exp{\left(\frac12u(z)+\mathrm{i}\theta(z)\right)},\label{c.18}\\
   &&A_1(z)=-2\mathrm{Re}\{\mathrm{i}\bar{\partial}\ln\phi\}, \quad A_1(z)=-2\mathrm{Im}\{\mathrm{i}\bar{\partial}\ln\phi\}.\label{c.19}
  \eerr
 Then $(\phi, A)$ is a  topological solution of the system
 \eqref{2.8}-\eqref{2.9}.

Hence the  proof of Theorem \ref{th3} is complete.

\section{ Existence of radially symmetric topological solutions and non-topological solutions }\label{s6}

\setcounter{equation}{0}
\setcounter{theorem}{0}\setcounter{lemma}{0}

 In this section we establish the existence
of  radially symmetric topological solutions and non-topological
solutions for  the generalized self-dual Chern--Simons equations
\eqref{2.8}-\eqref{2.9}, that is, we prove Theorem
\ref{th4}-\ref{th5}. We use the method developed in
\cite{chmy,yang1}.

  For convenience, we assume that the zero of $\phi$ concentrate
  at the origin with multiplicities $N$. Let $|\phi|^2=\re^u$, similar to Section
  \ref{s3}, we obtain the following governing equation
  \be
    \Delta  u=\lambda\mathrm{e}^u(\mathrm{e}^u-1)^5+ 4\pi N\delta(x),  \quad x\in\mathbb{R}^2.\label{5.1}
   \ee
Correspondingly, the topological condition \eqref{2.10} and
non-topological condition \eqref{2.11}  are changed as
   \be
   u(x)\to 0\quad\text{as}\quad |x|\to +\infty\label{5.2}
  \ee
and
 \be
   u(x)\to -\infty\quad\text{as}\quad |x|\to +\infty.\label{5.3}
  \ee

To get the proof of Theorem \ref{th4} and Theorem \ref{th5}, firstly
we  state the following theorem for \eqref{5.1}.

\begin{theorem}\label{th5.1}
 For $N\ge0$, a radially symmetric solution of \eqref{5.1} is either
 trivial $u\equiv0$, or negative $u<0$. For every given $N$ there
 exists a unique solution $u=u(r) (r=|x|)$ satisfies
 \[\lim\limits_{r\to+\infty}u(r)=0.\]
 All other solutions satisfies the the behavior
 \be\lim\limits_{r\to+\infty}ru_r(r)=-\beta, \quad \beta>2N+4.\label{5.4}\ee
  Furthermore, for any $\beta\in (2N+4, +\infty)$, there exists at
  least one solution $u$ of \eqref{5.1}  realizing the behavior
  \eqref{5.4}.
\end{theorem}

From the first part of Theorem \ref{th5.1}, we can get the existence
and uniqueness of radially symmetric topological solution, then  we
can get the proof of Theorem \ref{th4}.

 Let $u$ be a solution
obtained in the second part of Theorem \ref{th5.1}. Set
 \ber
  &&\phi(z)=\exp{\left(\frac12u(z)+\mathrm{i}N\arg z\right)},\label{5.5}\\
   &&A_1(z)=-2\mathrm{Re}\{\mathrm{i}\bar{\partial}\ln\phi\}, \quad A_2(z)=-2\mathrm{Im}\{\mathrm{i}\bar{\partial}\ln\phi\}.\label{5.6}
  \eer
  Then we can construct the
$N$-vortex radially non-topological solutions the generalized
Chern--Simons equations \eqref{2.8}-\eqref{2.9}. When
$\beta\in(2N+4, +\infty)$, let $u$ be a solution obtained in Theorem
\ref{5.1} such that \eqref{5.4} is satisfied.  Hence by
\eqref{2.8}-\eqref{2.9}, and \eqref{5.5}-\eqref{5.6}, we can obtain
the magnetic flux is
 \ber
  &&\Phi=\int_{\mathbb{R}^2} F_{12}\ud
  x=\frac\lambda2\int_{\mathbb{R}^2}\re^u(1-\re^u)^5\ud x\nonumber\\
  &&=\pi\int_0^{+\infty}(u_{rr}+\frac1ru_r)r\ud r=\pi\int_0^{+\infty}(ru_r)_r\ud r
  =\pi(2N+\beta).\label{5.7}
 \eer
Then the electric charge is \[Q=\kappa\Phi=\kappa\pi(2N+\beta).\]
Noting \eqref{5.4}-\eqref{5.5}  we can get
  \[|D_j\phi|^2=\frac12u_r^2\re^u=O(r^{-(2+\beta)}).\]
Then it follows from  \eqref{2.7'} that the energy is
 \[E=\int_{\mathbb{R}^2} \mathcal {E}\ud x=\int_{\mathbb{R}^2} F_{12}\ud x=\pi(2N+\beta).\]
Then we   complete the proof  of Theorem \ref{th5}.

 Now we just
need to prove Theorem \ref{th5.1}.

Since we are interested in radially symmetric solutions of
\eqref{5.1}, setting $r=|x|$, we obtain
 \ber
 &&u_{rr}(r)+\frac1ru_r(r)+\lambda\re^{u(r)}\big(1-\re^{u(r)}\big)^5=0, \quad r>0,\label{5.8}\\
 &&u(r)=2N\ln r+O(1), \quad  \text{for small  } r>0.\label{5.9}
  \eer
Using new variables
\[ t=\ln r,\quad u(t)\equiv u(\re^t),  \]
we transform \eqref{5.8}-\eqref{5.9} into
 \ber
  &&u''(t)+\lambda\re^{2t}\re^{u(t)}\big(1-\re^{u(t)}\big)^5=0, \quad -\infty<t<+\infty,\label{5.10}\\
 &&u(t)=2Nt+O(1), \quad  \text{as } t\to -\infty.\label{5.11}
 \eer

To prove Theorem \ref{th5.1}, we first  state the following theorem.

\begin{theorem}\label{th5.2}
 There exists a unique solution to \eqref{5.10}-\eqref{5.11} such
 that  $u\le 0, \, u'\ge0, \, u''<0$ in $\mathbb{R}$ and
 \be
 \lim\limits_{t\to+\infty}u(t)=0.\label{5.12}
 \ee

For any $\beta\in (2N+4, +\infty)$, the problem
\eqref{5.10}-\eqref{5.11} admits at least one solution such that
$u<0, \, u''<0$ in $\mathbb{R}$ and
 \be
  \lim\limits_{t\to+\infty}u'(t)=-\beta,\label{5.13}
 \ee
 Moreover, for any non-positive solution of
 \eqref{5.10}-\eqref{5.11} satisfying
  \be
 \liminf\limits_{t\to+\infty} u(t)<0,\label{5.14}
  \ee
  there exists some $\beta\in (2N+4, +\infty)$ such that $u$ satisfies
  \eqref{5.13}.
\end{theorem}

  Suppose that  $u$ is a solution of \eqref{5.10}-\eqref{5.11} which
  becomes positive at some $t=t_0$. Then it follows from maximum
  principle  that $u'(t_0)>0.$ Therefore, $u''(t)>0$ and $u'(t)>0$ for all
  $t>t_0$. Then we have
   \[ \re^{u(t)}-1>\re^{u(t_0)}-1>0  \quad \text{for all } t>t_0.\]
 Using the equation \eqref{5.10}, there exist a positive constant
 $\delta_0$ depending on $t_0$ such that
  \[u''(t)>\delta_0^4\re^{u(t)}, \quad t>t_0.\]
 Then it is easy to see that $u(t)$ blows up at  finite time
 $t>t_0$.

   Hence  by Theorem \ref{5.2}, we can conclude the assertion of Theorem \ref{5.1}.

  In the sequel we just need to Theorem \ref{5.2}.

 Let
   \berr
     g(u)=\left\{\begin{aligned}&\re^u(1-\re^u)^5,  & u\le0,\\
  &0, &u>0.
  \end{aligned}\right.
  \eerr
  To prove Theorem \ref{5.2}, it  is sufficient to prove the same result
  for the following problem
  \ber
  &&u''(t)+\lambda\re^{2t}g(u(t))=0, \quad -\infty<t<+\infty,\label{5.15}\\
 &&u(t)=2Nt+O(1), \quad  \text{as } t\to \infty.\label{5.16}
 \eer

 First we establish the existence for the initial value problem
 \eqref{5.15}.
 \begin{lemma}\label{lem5.1}
 For any $a\in \mathbb{R}$, there exits a unique solution $u$ to the
 problem \eqref{5.15} such that
 \be
 u(t)=2Nt+a+o(1) \quad  \text{ as}\quad  t\to-\infty.\label{5.17}
 \ee
 Moreover, if $u(t)$ is a solution of \eqref{5.15} in some interval,
  it can be extended to a global solution of \eqref{5.15} in
  $\mathbb{R}$ which satisfies  \eqref{5.17} for some $a\in
  \mathbb{R}$.
 \end{lemma}

  {\it Proof.} \quad It is easy to check that $u(t)$ is a solutio of
  \eqref{5.15} if and only if $u(t)$ verifies
   \be
  u(t)=2Nt+a-\lambda\int_{-\infty}^t(t-s)\re^{2s}g(u(s))\ud s.\label{5.18}
   \ee
  Let $T<-\ln2$, we can get
  \[ \int_{-\infty}^T(T-s)\re^{2s}\ud s<\frac{1}{16}.\]
 Noting that $|g(u)|+|g'(u)|< 7$, then by Picard iteration with
 $u^0=2Nt+a$, we can establish the solution of \eqref{5.15} in the
 interval $(-\infty, T]$.  Since $g(u)$ is bounded, we can extend
 $u$ to a solution of \eqref{5.15} in $\mathbb{R}$.

   Now we prove the uniqueness of the solution. Suppose that $u^1,
   u^2$ are two solutions of \eqref{5.15} in the interval $(-\infty, T]$. Let $\tilde{u}=u^1-u^2$, we have
   \berr
    |\tilde{u}(t)|&=& \left|-\lambda\int_{-\infty}^t(t-s)\re^{2s}\left[g(u^1(s))-g(u^2(s))\right]\ud s\right|\nonumber\\
    &\le&\sup|g'(u)|\int_{-\infty}^T(T-s)\re^{2s}\ud s \sup\limits_{(-\infty,\, T]}|\tilde{u}|\nonumber\\
     &\le&\frac12\sup\limits_{(-\infty,\, T]}|\tilde{u}|.
    \eerr
  Then we can get $\sup\limits_{(-\infty,\, T]}|\tilde{u}|=0$, which
  implies $u^1=u^2$ in $(-\infty, T]$.  By the unique continuation
  we have $u^1=u^2$ in $\mathbb{R}$.

   Now we prove the second part of the lemma. Assume that $u$ is a
   solution of \eqref{5.15} in some interval.  By
   \eqref{5.15}-\eqref{5.16} we have
    \[\lim\limits_{t\to-\infty}u'(t)=2N\]
    and
    \[u'(t)=u'(0)+\lambda\int_t^0\re^{2s}g(u(s))\ud s.\]
Then we have
 \[u'(t)=2N-\lambda\int_{-\infty}^t\re^{2s}g(u(s))\ud s.\]
Noting that
 \[\int_0^t\int_{-\infty}^s\re^{2s_1}g(u(s_1))\ud s_1\ud s=\lambda\int_{-\infty}^t(t-s)\re^{2s}g(u(s))\ud s<+\infty\]
we obtain
 \berr
  u(t)&=&2Nt+u(0)-\lambda\int_0^t\int_{-\infty}^s\re^{2s_1}g(u(s_1))\ud s_1\ud s\\
  &=&2Nt+\left(u(0)+\lambda\int_{-\infty}^0\int_{-\infty}^s\re^{2s_1}g(u(s_1))\ud s_1\ud s\right)
  -\lambda\int_{-\infty}^t\int_{-\infty}^s\re^{2s_1}g(u(s_1))\ud s_1\ud s
 \eerr
 That is, $u$ satisfies \eqref{5.17} with
 \[a=u(0)+\lambda\int_{-\infty}^0\int_{-\infty}^s\re^{2s_1}g(u(s_1))\ud s_1\ud s.\]
Then Lemma \ref{lem5.1} follows.

Now we investigate the behavior of the solutions as $t\to+\infty$.
In the sequel we denote by $u(t, a)$ by the solution given by Lemma
\ref{lem5.1}.  We use $'$ to denote the derivative with respect to
$t$ and subscript $_a$ to denote the derivative with respect to $a$.
We define the parameter sets:
 \berr
 \mathcal {A}^+&=&\{a\in\mathbb{R}|\quad \exists\, t_0\in\mathbb{R} \quad\text{such that}\quad u(t_0)>0\},\\
 \mathcal {A}^0&=&\{a\in\mathbb{R}|\quad  u(t, a)\le 0, \quad u'(t,a)\ge0, \,\, \forall t\in \mathbb{}R\},\\
 \mathcal {A}^-&=&\{a\in\mathbb{R}|\quad u(t,a)\le 0, \,\, \forall t\in\mathbb{R}, \quad \exists\, t_1\in\mathbb{R}\quad\text{such that}\quad u'(t_1)<0\}.
 \eerr

 It is easy to see that
  \[\mathcal {A}^+\cup\mathcal {A}^0\cup\mathcal {A}^-=\mathbb{R},\quad\mathcal {A}^+\cap\mathcal {A}^0
  =\mathcal {A}^0\cap\mathcal {A}^-=\mathcal {A}^+\cap\mathcal {A}^-=\emptyset\]
Furthermore, we can obtain the following lemma.
 \begin{lemma}\label{lem5.2}
  (1) If $a\in \mathcal {A}^+$, then $u'>0$ in the set $\{t|u(\tau,a)<0, \, \forall \tau\in(-\infty,
  t)\}$.\\
  (2) If  $a\in\mathcal {A}^0$, then $u''\le0$ and $u'\ge0$ in
  $\mathbb{R}$ and $\lim\limits_{t\to+\infty}u(t, a)=0$.\\
  (3) If $a\in\mathcal{A}^-$, then $u''<0$, $u<0$ in $\mathbb{R}$ and
  $\lim\limits_{t\to+\infty}u(t,a)=-\infty$.\\
  (4) $\mathcal {A}^+$ is open and $(\frac{5^5}{6^6}\frac{\lambda}{4}, +\infty)\subset\mathcal
  {A}^+$.\\
  (5) $\mathcal {A}^-$ is open. \\
  (6)  Let
  \[T>\frac12\ln\frac{2(2N+1)}{\lambda\re^{-1}(1-\re^{-1})^5(\re^{\frac1N}-1)},\]
  then $(-\infty, -\frac{5^5}{6^6}\frac{\lambda}{4}-2-2N)\subset\mathcal
  {A}^-$.\\
  (7)  $\mathcal {A}^0$ is nonempty, bounded and closed.
  \end{lemma}

  {\it Proof.}\quad (1) Let $a\in \mathcal {A}^+$ and $t_0$ be first time
  such that $u(t, a)$ hits the $t$ axis from below. Then, $u(t, a)<0$
  for all $t\in (-\infty, t_0)$. By the equation \eqref{5.15} we
  have $u''=-\lambda\re^{2t}g(u(t))<0$ in $(-\infty, t_0)$. Hence
  $ u'(t,a)>0$ in $(-\infty, t_0)$.

   (2) By the definition of $\mathcal {A}^0$,  we see that the limit
   $b\equiv\lim\limits_{t\to+\infty}u(t,a)$ exists and non-positive. If
  $b<0$, we have
  $\lim\limits_{t\to+\infty}u''(t,a)=-\lambda\lim\limits_{t\to+\infty}\re^{2t}\re^b(1-\re^b)^5=-\infty$,
  which leads to a contradiction.

   (3) If $u(t_0)=u'(t_0)=0$, then $u(t,a)\equiv0.$ Hence, if $a\in \mathcal
   {A}^-$, then $u(t,a)<0$ in $\mathbb{R}$, and $u''(t,a)<0$ by the
   equation \eqref{5.15}. Therefore,
   $\limsup\limits_{t\to+\infty}u'(t,a)<0$, which implies $\lim\limits_{t\to+\infty}u(t,a)=-\infty$.

  (4) Noting that $u(t,a)$ is continuous in $a$, if $u(t_0,a_0)>0$,
  then we have $u(t_0, a)>0$ when $a$ is in a small neighborhood of
  $a_0$. Hence $\mathcal {A}^+$ is open. By \eqref{5.18} we have
   \[u(0, a)=a-\lambda\int_{-\infty}^0\re^{2s}g(u(s))\ud s\ge a-\frac{5^5}{6^6}\frac{\lambda}{4}.\]
  If $a>\frac{5^5}{6^6}\frac{\lambda}{4}$, then $u(0, a)>0$,  which
  says  $a\in \mathcal {A}^+$.

  (5)  If $a_0\in \mathcal {A}^-$, then there exists
  $t_0\in\mathbb{R}$ such that $u'(t_0,a_0)<0$. Hence $u'(t_0,
  a)<0$ when $a$ is close to $a_0$. By (3) we have $u(t, a_0)<0$ for
  all $t\le t_0$ and $a$  close to $a_0$.   By \eqref{5.18}, we see
  that $u$ cannot take a local negative minimum. Then $u(t_0, a)<0$ and $u'(t_0, a)<0$ implies  $u'(t, a)\le 0$ for
  all $t>t_0$. Hence, $u(t, a)<0$ for all $t>t_0$ when a is close to
  $a_0$. Then we see that $\mathcal {A}^-$ is open.

  (6) Let $a<-\frac{5^5}{6^6}\frac{\lambda}{4}-2-2NT$. By
  \eqref{5.18} we have
  \[ u(t, a)\le 2Nt+a+\frac{5^5}{6^6}\frac{\lambda}{4}<-2 \quad \text{for all }\quad t\in(-\infty, 0].\]
 If $a\notin \mathcal {A}^-$, since that $u(t,a)$ cannot assume
 a local minimum, there exists constants $T_1$ and $T_2$ such that
 $T_2<T_1$, $u(t,a)\le -2$ in $(-\infty, T_2]$, $u(T_2, a)=-2, u'(T_2, a)\ge0,
 u(t,a)\in[-2,-1]$ for all $t\in[T_2, T_1]$, $u(T_1,a)=-1, u'(T_1,
 a)\ge0$. Then we get $u''(t,a)=-\lambda\re^{2t}g(u(t))\le 0$ for
 all $t\le T_1$, from which follows $u'(t,a)\le 2N$ for all $t\in (-\infty,
 T_1]$. Then we have $u(T_2, a)-u(0, a)\le 2NT_2$, which emplies
   \[T_2\ge \frac{u(T_2, a)-u(0, a)}{2N}\ge\frac{-2-a-\frac{5^5}{6^6}\frac{\lambda}{4}}{2N}>T.\]
Similarly, we have $T_1-T_2\ge \frac{1}{2N}$. Hence, by the choice
of $T$, we have
 \berr
 u'(T_1,a)&=&u'(T_2, a)-\lambda\int_{T_2}^{T_1}\re^{2s}\re^{u(s,a)}\big(1-\re^{u(s, a)}\big)^5\ud s\\
  &\le& 2N-\lambda\re^{-1}(1-\re^{-1})^5\int_{T_2}^{T_1}\re^{2s}\ud s\\
  &\le& 2N-\frac\lambda2\re^{-1}(1-\re^{-1})^5\re^{2T}(\re^{\frac1N}-1)<-1,
  \eerr
which leads to a contradiction. Therefore $a\in\mathcal {A}^-$.

(7) By the assertion of (4)-(6), we can get (7).

 Then the proof of
of Lemma \ref{lem5.2} is complete.

 Next we investigate the monotonicity of the solution with respect
 to $a$.
   \begin{lemma}\label{lem5.3}
   Let
   \[
   T_0(a)=\sup\left\{T\in [-\infty, +\infty]|\, u(t, a)<0, \, u'(t, a)>0,\, \forall t\in (-\infty, T]\right\}.
   \]
  Then,
  \[
   u_a(t, a)\ge\frac{1}{2N}u'(t, a)>0, \,\, \forall t\in(-\infty, T_0(a)).
   \]
   \end{lemma}

  {\it Proof.}\quad It is easy to see that $v(t, a)\equiv u_a(t, a)$
  exists, is smooth and verifies
    \ber
    && v''(t, a)=-\lambda\re^{2t}g'(u(t, a))v(t, a), \quad  -\infty<t<+\infty,\label{5.19}\\
     &&\lim\limits_{t\to-\infty}v(t, a)=1, \quad \lim\limits_{t\to-\infty}v'(t,  a)=0.\label{5.20}
    \eer
Let
 \[T_1(a)=\sup\{\tau\in\mathbb{R}|\, v(t, a)>0, \, t\in (-\infty, \tau)\}.\]
 Then by \eqref{5.20}, we see that $T_1(a)>-\infty$. Let $w(t, a)=u'(t,
 a)$, from \eqref{5.15} we have $\lim\limits_{t\to-\infty}w(t,a)=2N, \,\lim\limits_{t\to-\infty}w'(t,a)=0.$
 Hence the function $C(t, a)\equiv\frac{w(t,a)}{v(t,a)}, \, t\in(-\infty,
 T_1(a))$ satisfies $\lim\limits_{t\to-\infty}C(t,a)=2N, \,\lim\limits_{t\to-\infty}C'(t,a)=0.$
 Noting that $w(t, a)$ satisfies
  \[ w''(t, a)=-\lambda\re^{2t}g'(u(t,a))w(t, a)-2\lambda\re^{2t}g(u(t, a)),\]
then we obtain
 \[ C'(t, a)=-\frac{2\lambda}{v^2(t,a)}\int_{-\infty}^t\re^{2s}g(u(s, a))v(s,a)\ud s\le 0, \, \forall t\in(-\infty, T_1(a)).\]
Hence $C(t, a)\le 2N, \forall t\in(-\infty, T_1(a))$, which is
$u_a(t, a)\ge \frac{1}{2N}u'(t, a), \, \forall t\in(-\infty,
T_1(a))$. It is obvious that $T_1(a)\ge T_0(a)$, then  Lemma
\ref{lem5.3} follows.

Now we give a characterization of the sets $\mathcal{A}^+,
\mathcal{A}^-$, and $\mathcal{A}^0$.
 \begin{lemma}\label{lem5.4}
 There exists a constant $a_0$ such that
  \[\mathcal{A}^+=(a_0, +\infty), \quad \mathcal{A}^-=(-\infty, a_0), \quad\text{and }\quad \mathcal{A}^0=\{a_0\}.\]
 \end{lemma}
 {\it Proof.} \quad Step 1.  We show that there exists two constants
 $a_1,  a_2$ satisfying $a_1\le a_2$ such that
   \[\mathcal{A}^+=(a_2, +\infty), \quad \mathcal{A}^-=(-\infty, a_1), \quad\text{and }\quad \mathcal{A}^0=[a_1, a_2].\]
 To prove $\mathcal{A}^+=(a_2, +\infty)$, since $\mathcal{A}^+$ is
 open, we just need to prove that if $(b_1, b_2)\subset\mathcal{A}^+,$ then  $b_2\in \mathcal{A}^+$.
 For $a\in (b_1, b_2)$, let $z_0(a)$ be the first zero point of $u$
 on the $t$ axis. Then we have $u(z_0(a), a)=0,\,u'(z_0(a), a)>0$
 and by Lemma \ref{lem5.2} (1) $u'>0$ in $(-\infty, z_0(a)]$. Then
 by Lemma \ref{lem5.3} we obtain $u_a\ge\frac{1}{2N}u'>0$ in $(-\infty,
 z_0(a)]$.  Using the implicit function  theorem we see that
 $z_0(a)$ is differentiable with respect $a$ in the set $(b_1,
 b_2)$ and
  \[ \frac{\ud }{\ud a}z_0(a)=-\frac{u_a(t, a)}{u'(t, a)}<0.\]
In view of \eqref{5.18} we obtain $u(t, a)\le 2Nt+a$ in $(-\infty,
z_0(a)]$, which leads to $z_0(a)\ge -\frac{a}{2N}$  for $b\in (b_1,
b_2)$. Then we infer that $z_0(b_2)\equiv\lim\limits_{a\to
b_2}z_0(a)$ exists and is finite. From continuity, we have
$u(z_0(b_2), b_2)=0$. If $u'(z_0(b_2), b_2)=0$, then $u(t,
b_2)\equiv0$, which is impossible. Hence  $u'(z_0(b_2), b_2)\neq0$,
which concludes that $u(t, b_2)>0$ for $t$ near $z_0(b_2)$. Then
$b_2\in \mathcal{A}^+$.

To prove $\mathcal{A}^-=(-\infty, a_1)$, it is sufficient to prove
that   if $(b_1, b_2)\subset\mathcal{A}^-,$ then  $b_1\in
\mathcal{A}^-$.  For $a\in\mathcal{A}^-$, let $z_1(a)$ be the first
point such that $u'(z_1(a), a)=0$ and let $m(a)=u(z_1(a), a)$ be the
maximum of $u(\cdot, a)$ in $\mathbb{R}$. Noting that for $a\in
\mathcal{A}^-$, $u''(z_1(a), a)<0$, then again by the implicit
function theorem we see that $z_1(a)$ is a differentiable function
on $\mathcal{A}^-$. Hence we have
 \[ \frac{\ud m(a)}{\ud a}= u'(z_1(a), a)\frac{z_1(a)}{\ud a}+u_a(z_1(a), a)= u_a(z_1(a), a)\ge0, \, \forall a\in (b_1, b_2).\]
 Then we obtain
 \[m(a)=\sup\limits_{t\in \mathbb{R}} u(t, a)\le
 m\big(\frac12(b_1+b_2)\big)<0, \quad a\in \big(b_1, \frac12(b_1+b_2)\big).
 \]
Via continuity
 \[ m(b_1)=\sup\limits_{t\in \mathbb{R}} u(t, b_1)\le m\big(\frac12(b_1+b_2)\big)<0,\]
which implies $b_1\in \mathcal{A}^-\cup\mathcal{A}^0$.  From Lemma
\ref{lem5.2} (2), we see that $b_1\notin \mathcal{A}^0$. Then
$b_1\in \mathcal{A}^-$.

Noting that
$\mathcal{A}^0=\mathbb{R}\setminus(\mathcal{A}^-\cup\mathcal{A}^+)$,
then we have $\mathcal{A}^0=[a_1, a_2]$.

Step 2. We show that $a_1=a_2$. For $a\in \mathcal{A}^0$, we have
$u'(t,a)>0$ in $\mathbb{R}$ and by Lemma \ref{lem5.3} $u_a(t, a)>0$
in $\mathbb{R}$. Noting that $\lim\limits_{t\to+\infty}u(t, a)=0$,
for $\forall \delta>0$, there exists a continuous function
$T_\delta(a)$ such that $u(T_\delta(a))=-\delta$ and
$u(t,a)>-\delta$ in $(T_\delta(a), +\infty)$.  Since
$g'(u)=\re^u(1-\re^u)^4(1-6\re^u)\le 0$ when $u\in [-\ln6, 0]$, we
have
 \[ u''_a=-\lambda\re^{2t}\re^u(1-\re^u)^4(1-6\re^u)u_a\ge0, \quad  t\in[T_{\ln6}(a), +\infty), \,a\in[a_1, a_2].\]
 Therefore, $u_a$ is a non-negative convex function on $[T_{\ln6}(a),
 +\infty)$, $u_a(+\infty, a)\equiv\lim\limits_{t\to+\infty}u_a(t,
 a)$ exists and $u_a(+\infty, a)\in [0, +\infty]$.

  Now we show  $u_a(+\infty, a)>0 \,\,\forall a\in[a_1, a_2]$.  Suppose, on the contrary, $u_a(+\infty, a)=0$ for some $a
 \in [a_1, a_2]$. Then we see that $C(t,a)=\frac{u'(t,\, a)}{u_a(t,\, a)}$ satisfies
  \berr
  C'(t,a)&=&-\frac{2\lambda}{u^2_a(t,a)}\int_{-\infty}^t\re^{2s} \re^{u(s, a)}\big(1-\re^{u(s, a)}\big)^5u_a(s, a)\ud s\\
  &=&-\frac{2\lambda}{u^2_a(t,a)}\int_{-\infty}^0\re^{2s} \re^{u(s, a)}\big(1-\re^{u(s, a)}\big)^5u_a(s, a)\ud
  s \to -\infty\quad \text{as}\quad t\to +\infty,
  \eerr
 which concludes that $C(t, a)<0$ as $t$ is sufficiently large. This
 contradicts the fact that $C(t, a)>0$ for all $t\in \mathbb{R}$.
 Then we have   $u_a(+\infty, a)>0 \,\,\forall a\in[a_1, a_2]$.

 Then it follows from Fato's lemma that
  \[0=\lim\limits_{t\to+\infty}(u(t, a_2)-u(t, a_1))=\lim\limits_{t\to+\infty}\int_{a_1}^{a_2}u_a(t, a)\ud a\ge\int_{a_1}^{a_2}u_a(+\infty, a)\ud a,\]
 which implies $a_1=a_2$.

 The proof of Lemma \ref{lem5.4} is complete.

 \begin{lemma}\label{lem5.5}
 For  $a\in \mathcal{A}^-$, the limit
  \[\beta(a)\equiv-\lim\limits_{t\to+\infty}u'(t, a)\]
  exists and is positive and finite.
 \end{lemma}

{\it Proof.} \quad For $a\in \mathcal{A}^-$, $u''(t, a)<0$ in
$\mathbb{R}$, then $\beta(a)$ exists and $\beta(a)\in (0, +\infty]$.
We need to show  that $\beta(a)$ is finite.  Suppose that
$\beta(a)>3$, then there exits a constant $T>1$ such that $u(t,a)\le
-3t\quad \forall t\in(T, +\infty)$. When $u<-2$,
$g'(u)=\re^u(1-\re^u)^4(1-6\re^u)>0$. Then as $t>T$, we have $g(u(t,
a))=\re^{u(t, a)}\big(1-\re^{u(t, a)}\big)^5\le
\re^{-3t}\big(1-\re^{-3t}\big)^5$. Therefore, for all $t>T$
 \berr
  u'(t, a)&=&u'(T, a)-\lambda\int_T^t\re^{2s}\re^{u(s, a)}\big(1-\re^{u(s, a)}\big)^5\ud s\\
   &\ge& u'(T, a)-\lambda\int_T^t\re^{-s} \big(1-\re^{-3s}\big)^5\ud s\\
    &\ge& u'(T, a)-\lambda\int_0^{+\infty}\re^{-s} \big(1-\re^{-3s}\big)^5\ud s.
 \eerr
Then, we have $\beta(a)\le  u'(T, a)-\lambda\int_0^{+\infty}\re^{-s}
\big(1-\re^{-3s}\big)^5\ud s<+\infty$, which concludes the lemma.

\begin{lemma}\label{lem5.6}
 For $a\in \mathcal{A}^-$, the functions $\re^{2t} \re^{u(t, a)}\big(1-\re^{u(t, a)}\big)^5$ and $\re^{2t}\big[1-\big(1-\re^{u(t, a)}\big)^6\big]$ are both integrable
 on $\mathbb{R}$. Moreover,
  \ber
  \beta(a)+2N=\lambda\int_{\mathbb{R}}\re^{2t} \re^{u(t, a)}\big(1-\re^{u(t, a)}\big)^5\ud  t,\label{5.21}\\
  \frac{\beta^2}{2}-2N^2=\frac\lambda3\int_{\mathbb{R}}\re^{2t}\big[1-\big(1-\re^{u(t, a)}\big)^6\big]\ud  t.\label{5.22}
  \eer
\end{lemma}

 {\it Proof.} \quad  For $a\in \mathcal{A}^-$,  $u(t, a)<0$, then we have
 \[u'(t, a)=2N-\lambda\int_{-\infty}^t\re^{2s}\re^{u(s, a)}\big(1-\re^{u(s, a)}\big)^5 \ud s.\]
 In view of Lemma \ref{lem5.5}, we can take the limit $t\to+\infty$
 in the above expression to get
 \[-\beta(a)=2N-\lambda\int_{-\infty}^{+\infty}\re^{2s}\re^{u(s, a)}\big(1-\re^{u(s, a)}\big)^5 \ud s,\]
 which is \eqref{5.21}.

 Multiply \eqref{5.15} by $u'$, and integrating the resulting
 equation over $[-T, T]$, we have
 \ber
 \frac12u'^2(t,
 a)|_{t=-T}^{t=T}+\frac\lambda6\left\{\re^{2t}\big[1-\big(1-\re^{u(t, a)}\big)^6\big]\right\}\big|_{t=-T}^{t=T}
 =\frac\lambda3\int_{-T}^{T}\re^{2t}\big[1-\big(1-\re^{u(t, a)}\big)^6\big]\ud t.
 \label{5.23}
 \eer
 Since $u(t, a)<0 $ for $a\in \mathcal{A}^-, t\in\mathbb{R}$, it is
 easy to see that
  \be
  \lim\limits_{t\to-\infty}\frac\lambda6\left\{\re^{2t}\big[1-\big(1-\re^{u(t, a)}\big)^6\big]\right\}=0.\label{5.24}
  \ee
Noting that $u''(t, a)<0$ for $a\in \mathcal{A}^-, t\in\mathbb{R}$,
we have $u'(t, a)>-\beta(a)$, which implies $u'(t, a)>-\beta(a)$ for
$t\in \mathbb{R}$.  For $T$ sufficiently large such that $u'(t,
a)<0$ when $t>T$, we can obtain
 \berr
 \frac16\big[1-\big(1-\re^{u(t, a)}\big)^6\big]&=&\int_{-\infty}^{u(t, a)}\re^{u(s, a)}\big(1-\re^{u(s, a)}\big)^5\ud u(s, a)\\
 &=&\int_{+\infty}^t\re^{u(s, a)}\big(1-\re^{u(s, a)}\big)^5 u'(s, a)\ud s\\
 &\le& \beta(a)\int^{+\infty}_t\re^{u(s, a)}\big(1-\re^{u(s, a)}\big)^5\ud s
 \eerr
Then
 \be
  \frac\lambda6\re^{2t}\big[1-\big(1-\re^{u(t, a)}\big)^6\big]
  \le\lambda\beta(a)\int_t^{+\infty}\re^{2s}\re^{u(s, a)}\big(1-\re^{u(s, a)}\big)^5\ud s\to 0\quad \text{as}\quad t\to +\infty.\label{5.25}
 \ee
By \eqref{5.24} and \eqref{5.25},  letting  $T\to +\infty$  in
\eqref{5.23}, we  get \eqref{5.22}. Then Lemma \ref{lem5.6} follows.

 \begin{lemma}\label{lem5.7}
   $\beta(a)$ is continuous in $\mathcal{A}^-$.
 \end{lemma}
 {\it Proof.}\quad  For any $\tilde{a}\in\mathcal{A}^-$, we claim
 $\beta(\tilde{a})>2$. Suppose otherwise, noting that $u''<0$, we
 have $u'(t, \tilde{a})\ge-\beta(\tilde{a})>-2$ for $t\in
 \mathbb{R}$. Then we infer that there exists a positive constant
 $C$ such that $u(t, \tilde{a})\ge-C-2t$ for all $t>0$. Let $T$ be a
 time so that $u(t, \tilde{a})<-3$ for all $t>T$.  Then we have
 \berr
 \lambda\int_T^{+\infty}\re^{2t}\re^{u(t, \tilde{a})}\big(1-\re^{u(t, \tilde{a})}\big)^5\ud t
  &\ge& \lambda\int_T^{+\infty}\re^{2t}\re^{-C-2t}\big(1-\re^{-C-2t}\big)^5\ud t\\
  &=&\lambda\re^{-C}\int_T^{+\infty}\big(1-\re^{-C-2t}\big)^5\ud t=+\infty,
 \eerr
which contradicts the finiteness of $\beta(\tilde{a})$. Therefore
$\beta(\tilde{a})>2$.

Let $\delta=\frac{\beta(\tilde{a})-2}{4}$. Then there exists a
positive constant $T_1$ such that $T_1\ge \frac{3}{2+\delta}$,
 $u'(T_1, \tilde{a})<-2(1+\delta)$ and $u(T_1, \tilde{a})<-2(1+\delta)T_1$.
 Noting $u(t, a)$ and $u'(t, a)$ are both continuous with respect to
 $a$, we have $u'(T_1, a)<-(2+\delta)$ and $u(T_1,
 a)<-(2+\delta)T_1$ when a is close to $\tilde{a}$. Since $u''<0$,
 we have $u(t, a)\le -(2+\delta)t$ for $t\in [T_1, +\infty)$ when
 $a$ is close to $\tilde{a}$.   Let
 \berr
     w(t)=\left\{\begin{aligned}&\frac{5^5\lambda}{6^6}\re^{2t},  &\quad t\le T_1,\\
  &\lambda \re^{-\delta t}\big(1-\re^{-(2+\delta)t}\big)^5,&\quad t>T_1.
  \end{aligned}\right.
\eerr
 It is easy to see that $w(t)\in L^1(\mathbb{R})$. Noting that $\lambda\re^{2t}\re^{u(t, a)}(1-\re^{u(t, a)})^5\le
 w(t)$ for all $t\in \mathbb{R}$, then using Lebesgue dominated
 control theorem and \eqref{5.21}, we can get the continuity of
 $\beta(a)$.  The proof of Lemma \ref{5.7} is complete.

Now we want to obtain the range of $\beta(a)$.  We first investigate
the behavior of $\beta(a)$ as $a\to a_0$ and as $a\to -\infty$.
\begin{lemma}\label{lem5.8}
There holds the limit $\lim\limits_{a\to a_0}\beta(a)=+\infty.$
\end{lemma}
 {\it Proof.} \quad In view of \eqref{5.22}, the continuity of $u(t,
 a)$, and the fact $u(t, a_0)\le 0$, we can obtain
 \berr
 \liminf\limits_{a\to a_0}\frac12\beta^2(a)-2N^2&=&\liminf\limits_{a\to a_0}\frac\lambda3\int_{\mathbb{R}}\re^{2t}\big[1-\big(1-\re^{u(t, a)}\big)^6\big]\ud t\\
 &\ge& \liminf\limits_{T\to+\infty} \liminf\limits_{a\to a_0}\frac\lambda3\int_0^T\re^{2t}\big[1-\big(1-\re^{u(t, a)}\big)^6\big]\ud t\\
 &=&\lim\limits_{T\to+\infty}\frac\lambda3\int_0^T\re^{2t}\big[1-\big(1-\re^{u(t, a_0)}\big)^6\big]\ud t\\
 &\ge&\frac\lambda3\big[1-\big(1-\re^{u(0, a_0)}\big)^6\big]\lim\limits_{T\to+\infty}\int_0^T\re^{2t}\ud t=+\infty.
 \eerr
Noting that $\beta(a)>0$, we  conclude the lemma.

 \begin{lemma}\label{lem5.9}
  For $a\in \mathcal{A}^-$, let $m(a)=\sup\limits_{t\in\mathbb{R}}u(t, a)$.
  Then \[\lim\limits_{a\to-\infty}m(a)=-\infty.\]
 \end{lemma}

{\it Proof.}\quad Let $a\in \mathcal{A}^-$, and $z_1(a)$ be the
point such that $u'(z_1(a), a)=0$.  Then from \eqref{5.18} we have
$u(t, a)\le 2Nt+a$, which implies $m(a)=u(z_1(a), a)\le 2Nz_1(a)+a$,
that is $z_1(a)\ge \frac{m(a)-a}{2N}$.  Noting $0\le u'\le 2N$ in
$(-\infty, z_1(a))$, we have $m(a)-1<u(t, a)<m(a)$, for all $t\in
(z_1(a)-\frac{1}{2N}, z_1(a))$ such that
 \berr
 0=u'(z_1(a), a)&=&2N-\lambda\int_{-\infty}^{z_1(a)}\re^{2s}\re^{u(s, a)}\big(1-\re^{u(s, a)}\big)^5\ud s\\
  &\le& 2N-\inf\limits_{u\in [m(a)-1, m(a)]}\re^u(1-\re^u)^5\lambda\int_{z_1(a)-\frac{1}{2N}}^{z_1(a)}\re^{2s}\ud s.
 \eerr
Then we have
 \berr
  2N&\ge&\inf\limits_{u\in [m(a)-1, m(a)]}\re^u(1-\re^u)^5\lambda\int_{z_1(a)-\frac{1}{2N}}^{z_1(a)}\re^{2s}\ud  s\\
  &\ge&\inf\limits_{u\in [m(a)-1, m(a)]}\re^u(1-\re^u)^5\lambda\int_{\frac{m(a)-a-1}{2N}}^{\frac{m(a)-a}{2N}}\re^{2s}\ud s.
 \eerr
In view of $\re^{2s}\to +\infty$ as $s\to +\infty$, we we can obtain
$m(a)\to -\infty$ as $a\to-\infty$. Then Lemma \ref{lem5.9} follows.

\begin{lemma}\label{lem5.10}
 There holds the limit
 \be
 \lim\limits_{a\to-\infty}\beta(a)=2N+4.\label{5.25'}
 \ee
Moreover,
  \be
  \{\beta(a)|\, a\in\mathcal{A}^-\}=(2N+4,+\infty).\label{5.26'}
  \ee
\end{lemma}

{\it Proof.} \quad Let $a\ll-1$ and $T\gg1$ be two fixed constants.
Then by \eqref{5.22}, we have
 \berr
 \frac{\beta^2}{2}-2N^2&=&\frac\lambda3\int_{\mathbb{R}}\re^{2t}\big[1-\big(1-\re^{u(t, a)}\big)^6\big]\ud t\\
  &=&\frac\lambda3\int_{-\infty}^T\re^{2t}\big[1-\big(1-\re^{u(t, a)}\big)^6\big]\ud  t
  +\frac\lambda3\int_T^{\infty}\frac{\big[1-\big(1-\re^{u(t, a)}\big)^6\big]}{\re^{u(t, a)}\big(1-\re^{u(t, a)}\big)^5}\re^{2t} \re^{u(t, a)}\big(1-\re^{u(t, a)}\big)^5\ud t\\
  &=&\frac\lambda3\int_{-\infty}^T\re^{2t}\big[1-\big(1-\re^{u(t, a)}\big)^6\big]\ud t+G(T^*, a)\frac\lambda3 \int_T^{\infty}\re^{u(t, a)}\big(1-\re^{u(t, a)}\big)^5\ud t
 \eerr
where $T^*\in (T, +\infty)$ and
 \[G(T^*, a)=\frac{\big[1-\big(1-\re^{u(T^*, a)}\big)^6\big]}{\re^{u(T^*, a)}\big(1-\re^{u(T^*, a)}\big)^5}\]
Then in view of \eqref{5.21}, we  obtain
 \be
  \frac{\beta^2}{2}-2N^2=2G(T^*, a)(\beta(a)+2N)+H(T, a),\label{5.26}
 \ee
where
 \[ H(T, a)=\frac\lambda3\int_{-\infty}^T\re^{2t}\left\{\big[1-\big(1-\re^{u(t, a)}\big)^6\big]-G(T^*, a) \re^{u(t, a)}\big(1-\re^{u(t, a)}\big)^5\right\}\ud t.\]
 Using Lemma \ref{lem5.9}, we infer  that $\lim\limits_{a\to-\infty}H(T, a)=0$ and $\lim\limits_{a\to-\infty}G(T^*, a)=1$.
 It follows from \eqref{5.26} that
   \be
   \beta(a)=2G(T^*, a)+\sqrt{\big(2N+2G(T^*, a)\big)^2+2H(t, a)}.\label{5.27}
   \ee
Then, letting $a\to -\infty$ in \eqref{5.27}, we get \eqref{5.25'}.

 Since $\beta(a)$ is continuous in $\mathcal{A}^-$, we  obtain
 \be
  \{\beta(a)|\, a\in\mathcal{A}^-\}\subset(2N+4,+\infty).\label{5.28}
 \ee

Noting that as $a\in \mathcal{A}^-$, $u(t, a)<0,$ it is easy to
check that
 \[\frac{\frac16[1-(1-\re^{u(t, a)})^6]}{\re^{u(t, a)}\big(1-\re^{u(t, a)}\big)^5} > 1.\]
 Then in view of \eqref{5.22}, we have
   \berr
   \beta^2(a)-4N^2&=&\frac{2\lambda}{3}\int_{\mathbb{R}}\re^{2t}\big[1-\big(1-\re^{u(t, a)}\big)^6\big]\ud  t\\
   &=&4\lambda\int_{\mathbb{R}}\frac{\frac16\big[1-\big(1-\re^{u(t, a)}\big)^6\big]}{\re^{u(t, a)}\big(1-\re^{u(t, a)}\big)^5}\re^{2t}{\re^{u(t, a)}\big(1-\re^{u(t, a)}\big)^5}\ud  t\\
    &>&4\lambda\int_{\mathbb{R}}\re^{2t}{\re^{u(t, a)}\big(1-\re^{u(t, a)}\big)^5}\ud t\\
    &=&4(\beta(a)+2N),
   \eerr
 which implies $\beta(a)>2N+4$, that is
 \be
 \{\beta(a)|\, a\in\mathcal{A}^-\}\supset(2N+4,+\infty).\label{5.29}
 \ee
 Therefore \eqref{5.26'} follows from \eqref{5.28} and \eqref{5.29}.
 Then the proof of  Lemma \ref{lem5.10} is complete.

Now combining Lemmas \ref{lem5.1}-\ref{lem5.10}, we can get Theorem
\ref{th5.2}.

\end{document}